\documentclass[conference]{IEEEtran}

\usepackage{cite}
\usepackage{amsmath,amssymb,amsfonts}
\usepackage{algorithmic}
\usepackage{graphicx}
\usepackage{bmpsize}
\usepackage{textcomp}
\usepackage{xcolor}
\usepackage[hyphens]{url}
\usepackage{fancyhdr}
\usepackage[bookmarks=true,breaklinks=true,letterpaper=true,colorlinks,citecolor=blue,linkcolor=blue,urlcolor=blue]{hyperref}
\usepackage[normalem]{ulem}

\usepackage{multirow}
\usepackage{array}
\usepackage{textgreek}
\usepackage{soul}
\usepackage{balance}
\usepackage{makecell}
\usepackage{listings}
\def\BibTeX{{\rm B\kern-.05em{\sc i\kern-.025em b}\kern-.08em
    T\kern-.1667em\lower.7ex\hbox{E}\kern-.125emX}}

\newcommand{\textfw}[1]{\scalebox{.8}[1.0]{\texttt{#1}}}

\newcommand{\mmndp}{M$^2$NDP}  %
\newcommand{\mmfunc}{M$^2$func}  
\newcommand{\mmuthr}{M$^2\mu$thread}
\newcommand{\gpundpEq}{GPU-NDP(Iso-FLOPS)}
\newcommand{\gpundpQuad}{GPU-NDP(4$\times$FLOPS)}
\newcommand{\gpundpHexa}{GPU-NDP(16$\times$FLOPS)}
\newcommand{\gpundparea}{GPU-NDP(Iso-Area)}
\newcommand{\scndp}{NSU}

\title{Low-overhead General-purpose Near-Data Processing in CXL Memory Expanders} 
\author{
    \IEEEauthorblockN{
    Hyungkyu Ham\IEEEauthorrefmark{2}\IEEEauthorrefmark{1},
    Jeongmin Hong\IEEEauthorrefmark{2}\IEEEauthorrefmark{1}, 
    Geonwoo Park\IEEEauthorrefmark{2},
    Yunseon Shin\IEEEauthorrefmark{2},
    Okkyun Woo\IEEEauthorrefmark{2},\\
    Wonhyuk Yang\IEEEauthorrefmark{3},
    Jinhoon Bae\IEEEauthorrefmark{2}, 
    Eunhyeok Park\IEEEauthorrefmark{2}\IEEEauthorrefmark{3}, 
    Hyojin Sung\IEEEauthorrefmark{4},
    Euicheol Lim\IEEEauthorrefmark{6},
    Gwangsun Kim\IEEEauthorrefmark{2}}
    \and
    \thanks{\IEEEauthorrefmark{1} These authors contributed equally to this work.}
    \and
    \IEEEauthorblockA{\IEEEauthorrefmark{2}POSTECH\\Department of Computer\\ Science and Engineering}\and
    \IEEEauthorblockA{\hspace{0.15in} }\and
    \IEEEauthorblockA{\IEEEauthorrefmark{3}POSTECH\\ Graduate School of\\Artificial Intelligence}\and
    \IEEEauthorblockA{\hspace{0.15in} }\and
    \IEEEauthorblockA{\IEEEauthorrefmark{4}Seoul National University\\ Graduate School of\\Data Science}\and
    \IEEEauthorblockA{\hspace{0.15in} }\and
    \IEEEauthorblockA{\IEEEauthorrefmark{6}SK hynix}
    \and
    \IEEEauthorblockN{
        \IEEEauthorrefmark{2}\IEEEauthorrefmark{3}\{hhk971, jmhhh, geonwoo1998, ysshin, magel, wonhyuk, bae00003, eh.park, g.kim\}@postech.ac.kr
        \\
        \IEEEauthorrefmark{4}hyojin.sung@snu.ac.kr\hspace{1in}\IEEEauthorrefmark{6}euicheol.lim@sk.com}
    \vspace{-.1in}
}

\IEEEoverridecommandlockouts %

\begin{document}
\maketitle

\begin{abstract}
Emerging Compute Express Link (CXL) enables cost-efficient memory expansion beyond the local
DRAM of processors.
While its CXL.mem protocol provides minimal latency overhead through
an optimized protocol stack, frequent CXL memory accesses can result in 
significant slowdowns for memory-bound applications whether they are latency-sensitive or bandwidth-intensive.
The near-data processing (NDP) in the CXL controller promises to overcome 
such limitations of passive CXL memory. However, prior work on NDP in CXL memory
proposes application-specific units that are not suitable for practical
CXL memory-based systems that should support various applications. 
On the other hand, existing CPU or GPU cores are not cost-effective for NDP because they are not optimized for memory-bound applications. 
In addition, the communication between the host processor and CXL controller
for NDP offloading should achieve low latency, but existing CXL.io/PCIe-based mechanisms
incur $\mu$s-scale latency and are not suitable for fine-grained NDP.

To achieve high-performance NDP end-to-end, 
we propose a low-overhead general-purpose NDP architecture
for CXL memory referred to as \emph{Memory-Mapped NDP (M$^2$NDP)}, which 
comprises \emph{memory-mapped functions (\mmfunc{})} 
and \emph{memory-mapped $\mu$threading (\mmuthr{})}.
\mmfunc{} is a CXL.mem-compatible low-overhead communication mechanism
between the host processor and NDP controller in CXL memory.
\mmuthr{} enables low-cost, general-purpose NDP unit design 
by introducing lightweight \emph{$\mu$threads} that 
support highly concurrent execution of kernels with minimal resource wastage. 
Combining them, \mmndp{} achieves significant speedups for various workloads %
by up to 128x (14.5x overall) and reduces
energy by up to 87.9\% (80.3\% overall)
compared to baseline CPU/GPU hosts with passive CXL memory.
The \mmndp{} simulator is publicly available at \url{https://github.com/PSAL-POSTECH/M2NDP-public}.

\end{abstract}

\begin{IEEEkeywords}
Near-data procesing, CXL, Memory expander
\end{IEEEkeywords}

\section{Introduction}
\label{sec:intro}

The Compute Express Link (CXL)~\cite{cxl} interconnect standard is being widely adopted
for communication between processors, accelerators,
and memory expanders.
In particular, its \emph{memory-semantic} CXL.mem protocol enables
low-latency remote memory access with load/store instructions. 
The latency of CXL.mem is known to be significantly lower than that of PCIe and is
comparable to cross-socket NUMA latency, providing load-to-use
latency as low as 150-175~ns~\cite{cxl_scale, cxl_uiuc, pond, pcie-latency}. 
Thus, the host's memory capacity can be cost-effectively increased beyond the local DRAM,
benefiting workloads with huge memory footprints, 
including in-memory online analytic processing (OLAP), key-value store (KVStore), large language model (LLM)~\cite{gpt-3}, 
recommendation model (e.g., DLRM~\cite{dlrm}), and graph analytics~\cite{graph500}.

\begin{figure}
\vspace{0.10in}
\centering
\includegraphics[width=0.68\linewidth]{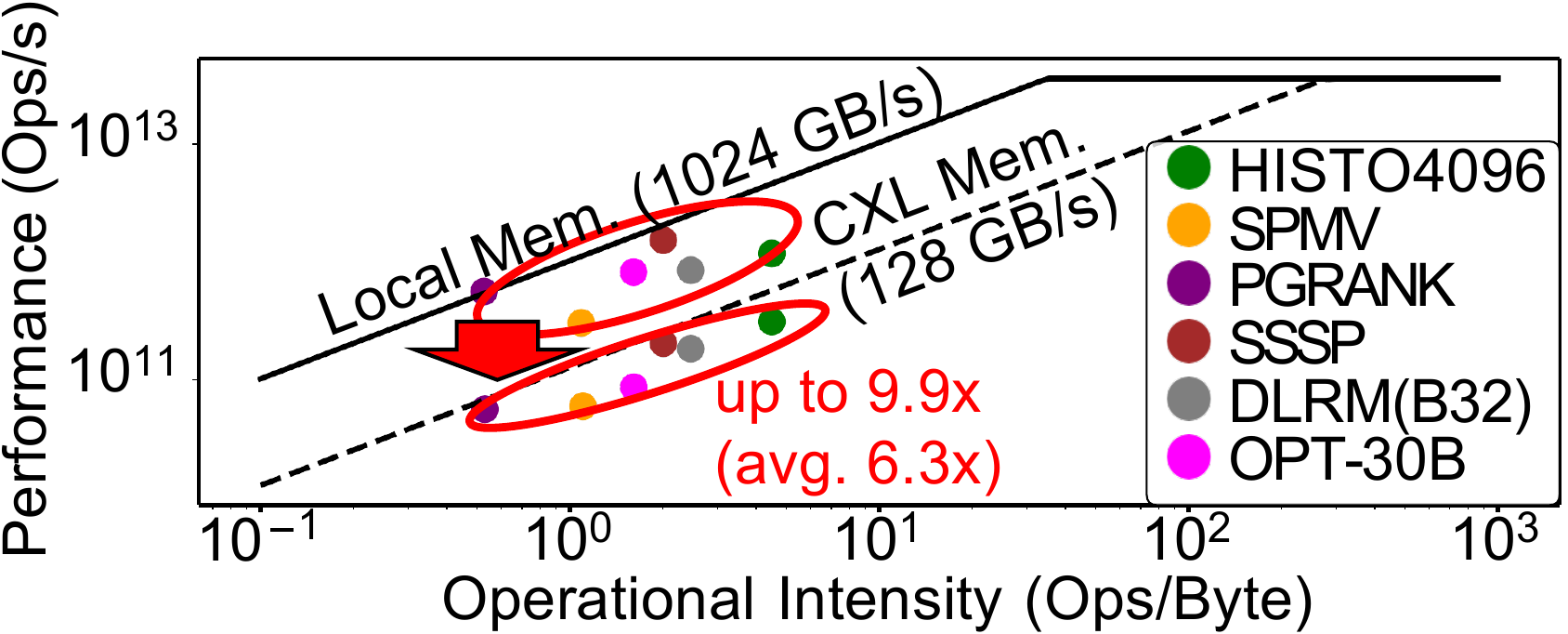}
\hspace{0.03in}
\includegraphics[width=0.26\linewidth]{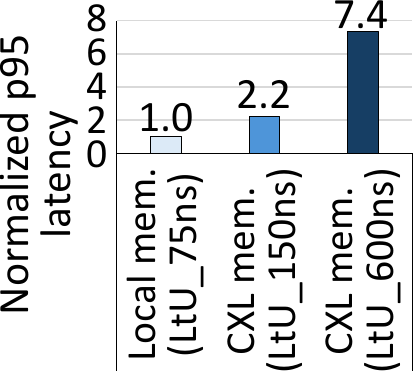}
\vspace{-.1in}
\hbox{\hspace{0.7in}\footnotesize{(a)} \hspace{1.7in} \footnotesize{(b)}\hspace{0.1in}}
\caption{{(a) Roofline analysis of workload performance with data 
in local memory vs. CXL memory. 
(b) Impact of Load-to-Use (LtU) latencies of local and CXL memories 
on the 95th percentile (P95) latency of key-value store (\textfw{KVS\_A}). 
CXL memory latency can vary depending on the implementation~\cite{cxl_uiuc, cmm-b, pond}.
Evaluation methodology is described in \S \ref{sec:methodology}.}}
 \vspace{-.2in}
\label{fig:impact_of_cxl_mem}
\end{figure}

However, the CXL link bandwidth (BW) can become a bottleneck for BW-intensive applications 
because it is substantially lower than the internal memory BW of 
CXL memories~\cite{cxl_cms_hynix_cal,cxl_cms_hynix_aicas}. 
As a result, placing the data of  applications that require both large memory capacity and high memory BW in CXL memory can substantially degrade performance by up to 9.9$\times$
(Fig.~\ref{fig:impact_of_cxl_mem}a).
CXL latency can also be significant for 
latency-sensitive applications that could exploit CXL memory to meet high memory capacity requirements (e.g., key-value stores) (Fig.~\ref{fig:impact_of_cxl_mem}b)~\cite{cxl_uiuc, tpp, pond}.
To address these limitations of passive CXL memory, several prior works
propose accelerating memory-bound workloads
with near-data processing (NDP) in CXL 
memory~\cite{cxl_anns, beacon, cxl_cms_hynix_aicas}.

Unfortunately, these prior works propose domain-specific NDP HW 
logic in CXL memory, limiting their target workloads.
Moreover, adding a wide variety of special-purpose NDP HW for different
NDP targets in CXL memory may not be a practical approach due to the
high total area and NRE costs~\cite{asic_clouds}. 
While FPGAs can adapt to target workloads~\cite{cxl_fpga}, 
they pose considerable programmability challenges~\cite{fpga_programmability}.
Meanwhile, for memory-bound workloads with little data reuse, general-purpose NDP can achieve performance similar to specialized logic as long as the memory bandwidth is saturated.
However,
existing CPU or GPU cores, when used as NDP units~\cite{ndp_sc,ndc,
top_pim, abndp, mobile_ndp, ndp_pact15, mondrian}, do not
provide sufficient performance per cost based on our evaluation,
because they are not optimized for memory-bound workloads.

Furthermore, 
conventional ring buffer and MMIO-based NDP offloading using CXL.io/PCIe~\cite{cxl_anns, beacon, cxl_cms_hynix_aicas, cxl_cms_hynix_cal}
can incur high latency due to the complexity of the scheme,
the CXL.io protocol stack overhead, and costly kernel mode switching on the host, which wastes CPU cycles.
While CXL.mem has low latency and can be used within user space,
it only supports basic memory reads/writes. 
Therefore, for latency-sensitive fine-grained NDP,
low-overhead offloading mechanism is necessary.

To this end, we propose a novel \emph{Memory-Mapped NDP (\mmndp{})} architecture to realize low-overhead, general-purpose NDP in CXL memory. 
\mmndp{} is based on two key components we propose: 
\emph{Memory-Mapped function (\mmfunc{})} for low-overhead communication 
between the host and NDP-enabled CXL memory, and \emph{Memory-Mapped $\mu$threading
(\mmuthr{})} for efficient NDP kernel execution.

The \mmfunc{} selectively repurposes CXL.mem packets 
for efficient host-device communication in NDP. %
By encapsulating NDP management commands (i.e., function calls) in CXL.mem 
requests to pre-determined addresses, we can avoid the high latency overhead of
conventional offloading using CXL.io/PCIe. 
A key enabler for the \mmfunc{} is a \emph{packet filter} placed at the input port
of the CXL memory. It checks if an incoming request's memory address
matches the pre-allocated memory range dedicated for each host process.
Then, for matching requests, different NDP management functions are triggered 
depending on the address. Thus, NDP management function calls (e.g., kernel registration, launch, and 
status poll) can be done simply by issuing memory accesses from the host. 
As a result, \mmfunc{} minimizes the latency of NDP offloading, especially benefiting 
fine-grained NDP. Additionally, we do not require any modification to the CXL.mem 
standard for best compatibility with host CPUs.
Consequently, \mmfunc{} avoids the complexity of managing a ring buffer-based shared task queue between the host and CXL/PCIe-attached devices by providing a clean function call abstraction. 

Furthermore, we propose \mmuthr{}
for the intuitive abstraction of NDP and cost-effective kernel execution.
Memory-bound workloads tend to use fewer registers than compute-bound workloads.
Thus, we propose a \emph{$\mu$thread}, which is a lightweight thread with a subset of the architectural
registers, as a unit of execution. 
By reducing register usage, the NDP unit can concurrently execute many $\mu$threads 
with fine-grained multithreading (FGMT)
to hide DRAM access latency without excessive physical register file cost.

In addition, memory-bound data-parallel workloads are typically implemented such that 
each thread is associated with specific data to be processed. 
In conventional programming environments such as CUDA, 
the association between a thread and memory location is expressed 
\emph{indirectly} via code (e.g., calculating the index of the array element for a thread using
threadblock ID, block dimension, and thread ID in CUDA).
In contrast, with our \mmuthr{}, each $\mu$thread is created in \emph{direct}
association with a particular memory location -- i.e., the $\mu$threads are memory-mapped, thereby reducing the code needed for address calculation.
The mapping is done at a smaller granularity (e.g., 32 B) compared to CUDA, which typically maps 128 B of data to a warp,
in order to mitigate the impact of intra-warp divergence.
Furthermore, to avoid the redundant
address calculation in SIMT-only GPUs~\cite{r2d2}, scalar instructions are supported.
Our NDP unit adopts a modified RISC-V ISA with 
vector extension (RVV)~\cite{risc-v-vector} to leverage SIMD units 
and fully utilize the DRAM BW within CXL memory cost-effectively.
The $\mu$threads are spawned individually, unlike thread block spawning in GPUs, which can waste resources due to inter-warp divergence.
Our on-chip scratchpad memory with a broader
scope than that of GPUs also reduces memory traffic 
and synchronization.

Overall, our proposed \mmndp{} architecture enables low-overhead, general-purpose NDP in CXL memory.
We show the effectiveness of our design for various workloads, 
including in-memory OLAP, KVStore, LLM, DLRM, and graph analytics. %

To summarize, our contributions include the following:

\begin{itemize}
    \item We propose \emph{\mmndp{} (memory-mapped NDP)} to enable
    \textbf{general-purpose} NDP in CXL memory.
    Our architecture is based on the unmodified CXL.mem protocol and, thus, does not require any modifications to the host processor hardware.
    \mmndp{} consists of \emph{\mmfunc{} (memory-mapped function)} and \emph{\mmuthr{} (memory-mapped $\mu$threading)}.

    \item The \mmfunc{} supports \textbf{low-overhead NDP offloading
    and management} from the host processor through CXL.mem, 
    overcoming the high overhead of conventional CXL.io/PCIe-based
    schemes for fine-grained NDP offloading
    while retaining standard-compatibility. As a result, it achieves speedups of up to 
    2.38$\times$ (16.8\% overall) compared to NDP offloading with 
    conventional approaches.

    \item The \mmuthr{} with lightweight FGMT based on a modified RISC-V with vector extension can    
    \textbf{cost-efficiently execute NDP kernels}.
    Compared to GPU cores, it reduces waste of 
    on-chip resources and
    memory bandwidth due to redundant address
    calculation, inter/intra-warp divergence, and 
    limited on-chip scratchpad memory scope.

    \item \mmndp{} can achieve high speedups
    of up to 128$\times$ (14.5$\times$ overall) for various workloads, compared to
    the baseline system with passive CXL memory, while reducing energy consumption
    by up to 87.9\% (80.3\% overall).

\end{itemize}

\section{Background and Motivation}

\subsection{Considerations in Architecting NDP in CXL Memory}
\label{sec:target}

While passive CXL memory can degrade the performance of latency- and BW-sensitive 
workloads~\cite{cxl_uiuc, astraea}, 
NDP in CXL memory presents a substantial opportunity to effectively address this challenge. 
Although NDP in CXL memory offloads host computation similar to GPUs,
they are introduced with very different \emph{primary} objectives 
(i.e., memory expansion vs. compute acceleration) and, thus, have fundamentally different
requirements for memory capacity, cost, and compute throughput (Table~\ref{tab:gpu_comp}).
In particular, CXL memory cannot have 100s of SMs as in high-cost GPUs~\cite{a100}.
The NDP also specifically targets
memory-bound %
workloads with low arithmetic intensity and large memory footprints
that do not fit in on-chip caches; 
other workloads (compute-bound or small working set) %
can be executed more efficiently on the host or GPUs.

\begin{table}
\scriptsize
    \centering
    \caption{High-level comparison of GPU and CXL memory with NDP.}
    \begin{tabular}{|>{}l||>{}c|>{}c|}
    \hline 
         & {\bf GPU} & {\bf CXL memory with NDP} \\ \hline \hline
         {\bf Memory capacity}  & Low & High \\ \hline
         {\bf Cost (area and power)} & High & Low \\ \hline
         {\bf FLOPS per memory BW} & High & Low \\  \hline
         {\bf Key target workloads} & Compute-bound & Memory-bound \\ \hline
    \end{tabular}
    \label{tab:gpu_comp}
\end{table}

\subsection{Compute Express Link Interconnect}
\label{sec:cxl_background}

\begin{figure}
\centering
\includegraphics[width=1.00\linewidth]{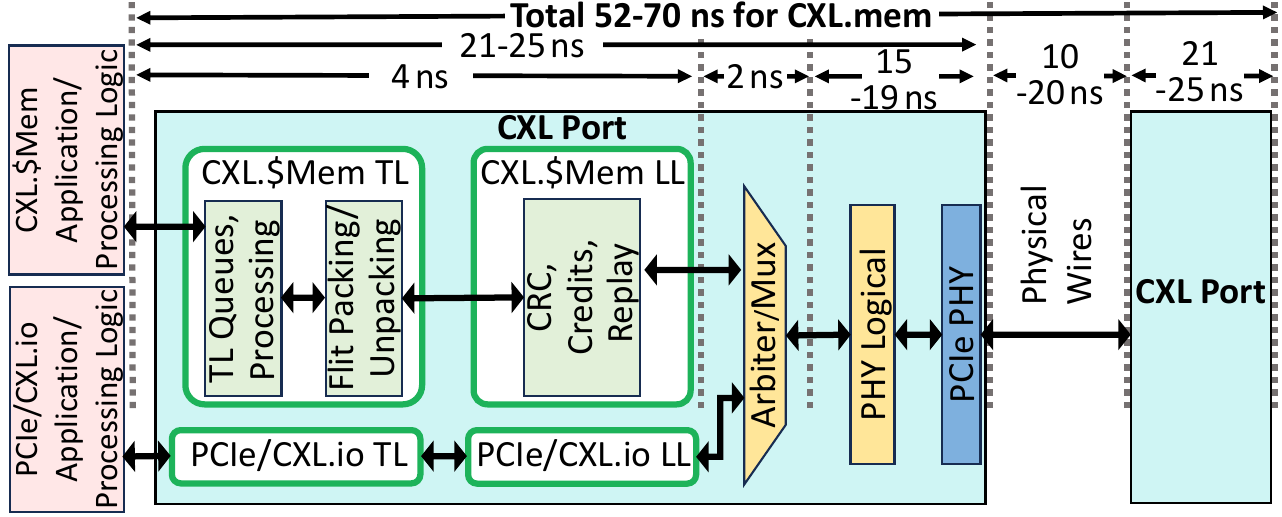}
\caption{CXL implementation and measured \emph{round-trip} latencies for CXL.mem (figure and numbers adapted from D. D. Sharma~\cite{cxl_scale}). CXL.\$Mem refers to both CXL.cache and CXL.mem (TL: transaction layer, LL: link layer).}
\label{fig:cxl_path}
\end{figure}

CXL~\cite{cxl} uses PCIe's PHY layer and defines three protocols: 
CXL.io (equivalent to PCIe) for device management;
CXL.cache for cache coherence between the host and device; %
CXL.mem for memory expansion through CXL.
In particular, CXL.mem %
enables processors 
to access CXL memory data by simply
issuing load/store instructions while providing lower latency compared to CXL.io/PCIe~\cite{tpp, cxl_scale, pcie-latency}.
The load-to-use latency for CXL memory can be as low as 
$\sim$150~ns, which includes \emph{round-trip} latencies through the host cache, CXL protocol
stack, physical off-chip wires, and DRAM~\cite{cxl_scale,pond, cxl_intro}. A round-trip latency through the CXL protocol stack and physical wires 
can be done in $\sim$70~ns (Fig.~\ref{fig:cxl_path}). 
When a CXL switch is used, the CXL memory access 
latency can approach 300~ns~\cite{pond}.
In contrast, CXL.io/PCIe takes $\sim$1$\mu$s or higher latency for communication (\S \ref{sec:pcie_overhead}).
The CXL.io is required for all CXL devices for device management.

The host manages the CXL memory, referred to as Host-managed Device Memory (HDM), which can be accessed using a Host Physical Address (HPA).
The HDM can use either HDM-H (host-only coherent) 
or HDM-DB (device coherent using back-invalidation) model.
The HDM-H is for passive memory expanders that do not 
manipulate the memory exposed to the host~\cite{cxl}.
In contrast, HDM-DB supports a device coherence agent (DCOH) and a snoop filter 
in CXL memory to track the host's caching of HDM, %
so it can back-invalidate (BI) the host cache using BI channels of CXL.mem when needed~\cite{cxl}.
Thus, HDM-DB is suitable
for CXL memory with NDP capability, and we use it
for \mmndp{}. 
The host can also flush HDM data from its cache using 
HW support~\cite{arm-hw-flush, intel_dsa, intel-opt-manual}.

The CXL 3.0 also supports direct peer-to-peer (P2P) access, allowing a CXL device to 
directly access the HDM of another CXL device through a CXL switch~\cite{cxl_p2p}. %
It can be useful for scalable NDP across multiple CXL memories.
However, accessing host memory from a CXL device is not supported by CXL.mem.
To access its own memory, a CXL device can use the Address Translation Services (ATS)~\cite{ats}
defined in PCIe to request a translation by the host, but it 
can incur a $\mu$s-scale latency due to protocol overhead 
and page table walks on the host~\cite{ats_latency}.
To reduce the overhead, the device can have an Address Translation Cache (ATC)
to keep recently used translation information.
When necessary, the host can invalidate the device's ATC to 
prevent incorrect translations.

\subsection{Communication Overhead with CXL.io/PCIe}
\label{sec:pcie_overhead}

Computation offloading with CXL.io/PCIe 
involves several SW and HW steps with significant overhead 
in terms of latency and host processor usage, especially for 
fine-grained offloading.
A common method used for GPUs and IO devices is based on
a ring buffer shared and manipulated by both the host driver and a PCIe device~\cite{ring_buffer}. 
For example, to launch a GPU kernel, the host runtime first writes the kernel launch command 
in the user buffer and the driver pushes a packet that points to the GPU command 
into the ring buffer in the kernel space. 
The host then updates the write (or tail) pointer of the ring buffer to notify the GPU
of the new command~\cite{amd_ring_buffer, ring_buffer2}, which incurs additional latency through PCIe
and triggers two DMA operations from the GPU to fetch the GPU command. 
Overall, the complex manipulation of the ring buffer shared between the
host and GPU can incur two and a half CXL.io round-trips for a kernel launch~\cite{ring_buffer}, 
resulting in high latency of $\sim$3-6$\mu$s~\cite{fe-bit, cuda-launch-overhead} (\S \ref{sec:ndp_launch}).
To check kernel completion, polling or interrupt is done, but 
polling over PCIe can require 2-3$\mu$s~\cite{ark}, and interrupt has
similar or higher overhead~\cite{interrupt_delay, poll_better, poll_energy}.
DMA over PCIe also incurs at least $\sim$1$\mu$s latency~\cite{pcie-latency}.
Thus, the latencies of kernel launch and completion check can be significant, especially 
for latency-sensitive, fine-grained NDP. 

Alternatively, to avoid such overhead,
a pair of device-side registers can be directly accessed through MMIO
over PCIe to send a request and check the result~\cite{nda, cxl_cms_hynix_cal, cxl_cms_hynix_aicas}.
However, it cannot support multiple
concurrent requests, resulting in limited throughput.
In addition, since the memory-mapped registers are physical resources, they cannot be safely shared among multiple
user processes and require a context switch to kernel space for every access.

\section{Memory-Mapped Near-data Processing}

\subsection{Overview}
\label{sec:overview}

\begin{figure}
\centering
\includegraphics[width=0.80\linewidth]{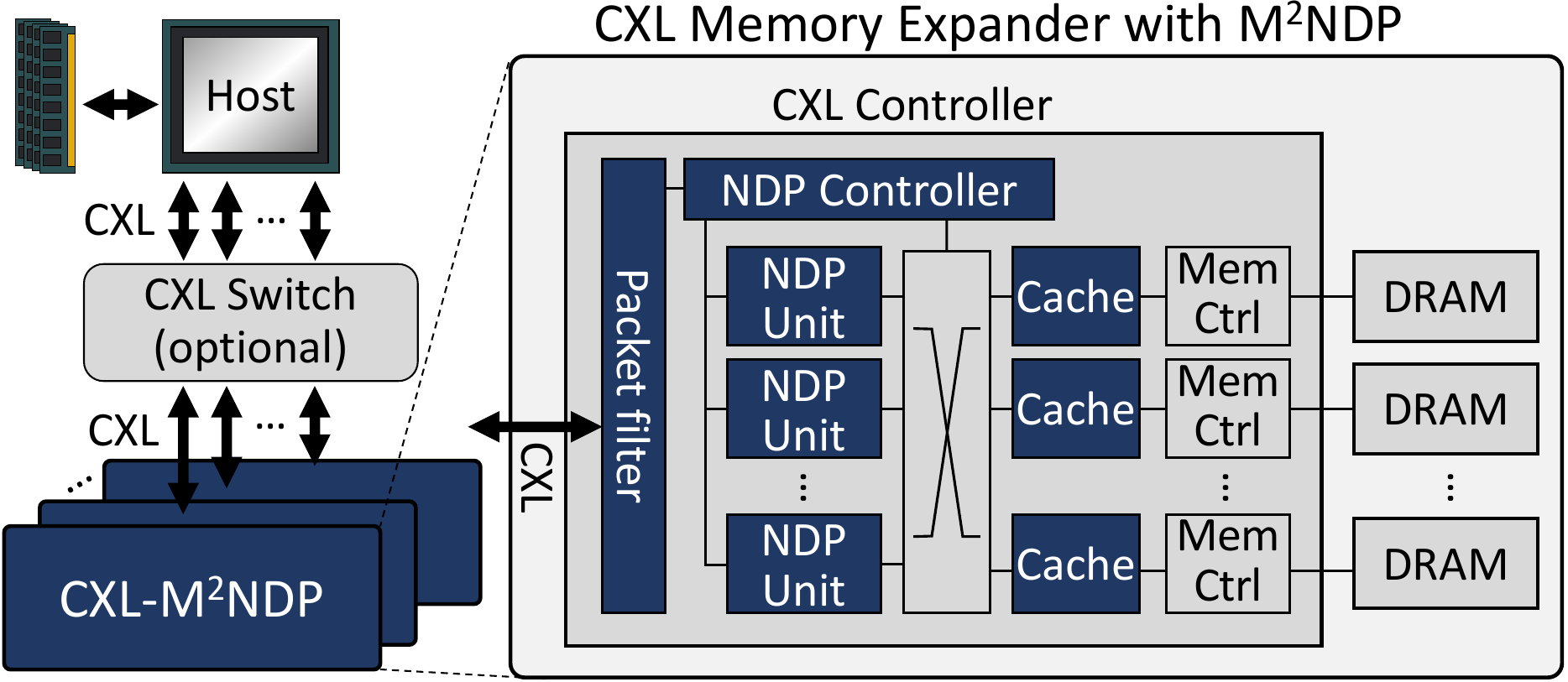}
\caption{Overview of the proposed system with \mmndp{}-enabled CXL memory.}
\label{fig:overview}
\end{figure}

To overcome the limited flexibility and cost-efficiency of 
prior NDP approaches while avoiding the high latency overhead 
in the offloading procedure (\S \ref{sec:pcie_overhead}) for NDP in CXL memory, we propose
\emph{Memory-Mapped Near-Data Processing (\mmndp{})} 
in CXL memory, called CXL-\mmndp{} (Fig.~\ref{fig:overview}).
The \mmndp{} comprises two mechanisms -- 1) \emph{Memory-Mapped functions (\mmfunc{})} for
low-overhead NDP management and offloading based on unmodified standard CXL.mem and 
2) \emph{Memory-Mapped $\mu$threading (\mmuthr{})}
for cost-effective general-purpose NDP microarchitecture.
They are combined to holistically improve end-to-end NDP performance 
including both offloading procedures and kernel execution.
They are implemented in the CXL controller chip which 
also supports the basic read/write CXL.mem transactions.

\subsection{Memory-mapped NDP Management Function (\mmfunc{})} 
\label{sec:management}

To exploit NDP for fine-grained computation offloading as well as coarse-grained
offloading, the communication latency between the host and CXL-\mmndp{} needs to be
minimized. While the CXL.mem protocol provides low latency, the standard
only defines packet types for normal CXL memory accesses and cannot be directly 
used for other communication. 
To extend CXL.mem to support custom packet types,
the host processor HW should be modified to support the special usage of the
reserved bits. Thus, commodity processors that only support the standard 
protocol cannot utilize it. 
Furthermore, to send special packets, special instructions would need to be introduced
in the host's ISA as in prior works~\cite{opportunistic_gpu, ndp_sc, tom}.
Such propriety extension of the standard protocol or host's ISA would hinder widespread adoption.
In contrast to CXL.mem, the conventional PCIe/CXL.io-based ring buffer scheme
supports arbitrary communication, but incurs higher latency from the protocol 
stack, ring buffer management, and context switch to the OS for privileged IO device communication (\S\ref{sec:pcie_overhead}).

Thus, to enable low-overhead and flexible communication with CXL-\mmndp{} from the host
using unmodified CXL.mem, we propose \mmfunc{}.
The basic idea is to reserve some physical memory space of the CXL memory,
referred to as the \emph{\mmfunc{} region}, 
for host communication.
To distinguish between the two different usages of CXL.mem, we introduce a \emph{packet filter} 
placed at CXL memory's input port to examine all packets
and determine if the packet should be interpreted as normal reads/writes
or \mmfunc{} call based on the packet's address. 
\mmfunc{} can be used to provide various functionalities, 
including NDP kernel registration/unregistration and launch.
Different functions can be called by using corresponding offsets 
from the base of the \mmfunc{} region for the CXL.mem packet (Table~\ref{tab:ndp_func}).
\mmfunc{} calls are handled by the NDP controller (Fig.~\ref{fig:overview}), which is 
implemented similarly to the microcontrollers in GPUs~\cite{nvidia-risc-v}.

\begin{figure}
\centering
\includegraphics[width=0.95\linewidth]{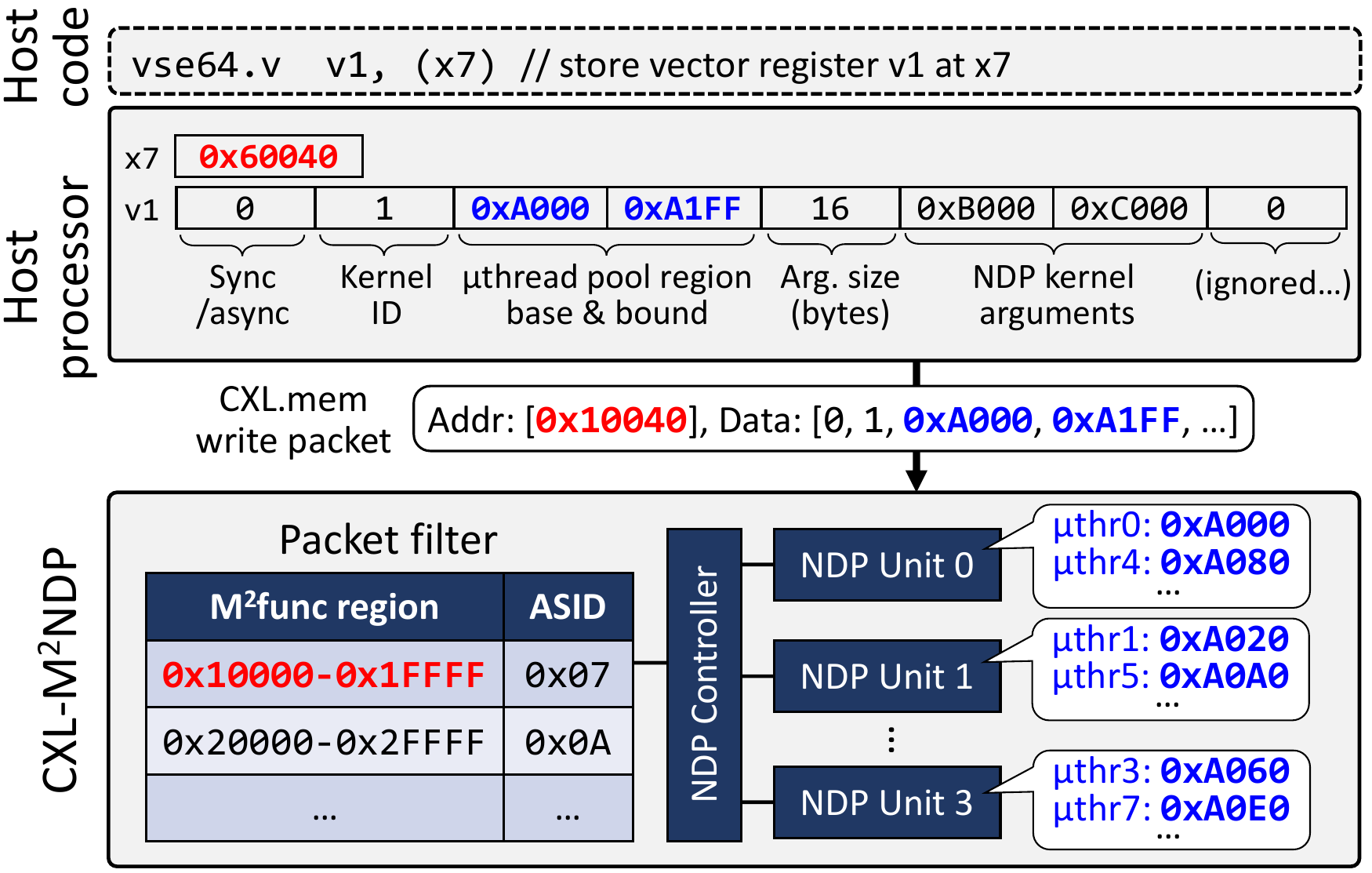}
\caption{Example NDP kernel launch using \mmfunc{} with VectorAdd NDP kernel that computes C=A+B. 
Vectors A, B, and C are placed at 0xA000, 0xB000, and 0xC000, respectively. 
It is assumed that the virtual address 0x60040 is translated into physical address 0x10040.
Each $\mu$thread computes a 32B (8x4B) partial vector output. 
Other datapath components are not shown.
}
\label{fig:ndp_example}
\end{figure}

For the initialization of \mmfunc{}, a host's user process can
request the \mmndp{} driver to allocate  
an uncacheable \mmfunc{} region in CXL memory
and insert its physical address range into 
the packet filter using the CXL.io scheme.
Once initialized, CXL.io is not needed anymore for NDP and
CXL.mem can be used for both normal reads/writes and \mmfunc{}. 

The packet filter entry requires little storage of only 
18~B per host process (64-bit base, 64-bit bound, and 16-bit ASID), so a small packet
filter can support many processes (e.g., 18~KB for 1024 processes)
and can also be easily 
replicated in multi-ported CXL memory~\cite{cxl}.

\begin{table*}[]
\centering
\scriptsize
\caption{Proposed user-level library API for \mmndp{}. \textfw{ERR} is a negative value representing an error.}
\setlength{\tabcolsep}{1.5pt} %
\begin{tabular}{|l|l|l||c|c|}
\hline
{\bf API Function} & {\bf Arguments} & {\bf Return Value} & {\bf Privileged }& {\bf Offset} \\
\hline \hline
ndpRegisterKernel & codeLoc, scratchpadMemSize, numIntRegs, numFloatRegs, numVectorRegs  & ndpKernelID or \textfw{ERR} & No  & 0  \\ \hline
ndpUnregisterKernel & ndpKernelID & 0 (success) or \textfw{ERR} & No & 1 $\ll$ 5 \\ \hline
ndpLaunchKernel & synchronicity, ndpKernelID, $\mu$threadPoolRegion (base, bound), kernelArgSize, kernelArguments & kernelInstanceID or \textfw{ERR} & No  & 2 $\ll$ 5 \\ \hline
ndpPollKernelStatus & ndpKernelInstanceID & 0 (finished), 1 (running) & No & 3 $\ll$ 5 \\ 
    & & 2 (pending), or \textfw{ERR} & & \\ \hline
ndpShootdownTlbEntry & ASID, virtualPageNumber & 0 (success) or \textfw{ERR} & Yes & 4 $\ll$ 5 \\ \hline
\end{tabular}
\label{tab:ndp_func}
\end{table*}

For an \mmfunc{} call, we use a write request format to include
arguments in the write data portion of the request.
To send it, the host executes a store instruction with a register
that holds the arguments (Fig.~\ref{fig:ndp_example}),
which will bypass the host cache as 
the \mmfunc{} region is uncacheable. 
Vector registers~\cite{sve, risc-v-vector, avx} 
can be used to send multiple arguments up to the vector register's size. 
However, the response to the write request cannot include any return value data 
from the NDP controller using the CXL.mem. 
Thus, we use a subsequent read request to the same address to access the return value
of the latest call of the function by the current process. 
Because the return value will be accessed with a normal memory access, 
the NDP controller can simply store the function's return value at the corresponding
memory address and serve the read request as normal access. 
For proper ordering, the host process code should have a fence instruction between 
the requests.

Table~\ref{tab:ndp_func} lists the NDP management functions for different address offsets
from the base of an \mmfunc{} region. 
To support sufficient sizes for function arguments and return values, 
the offsets can be strided (by 1$\ll$5 or 32~B in this example). Thus, multiple arguments and return values can be communicated.
For example, to register (unregister) an NDP kernel,
assuming the base address is \textfw{0x00FF0000}, a write request to  
\textfw{0x00FF0000} (\textfw{0x00FF0020}  or  \textfw{0x00FF0000+(1$\ll$5})) can be used.
Since different kernels can require 
varying amounts of register and scratchpad memory (\S\ref{sec:ndp_kernel}),  
they are given as arguments for registration.
In addition, the kernel argument size should be specified so that the arguments
can be properly extracted from a kernel launch packet. 
The metadata of registered kernels are stored in the \mmfunc{} region for the
current host process, beginning at a pre-determined location 
beyond the offsets used in Table~\ref{tab:ndp_func} for ease of accesses by the host.
As the \mmfunc{} region is allocated by each process, it is protected from other processes by the host.

\subsection{NDP Kernel Launch}
\label{sec:ndp_launch}

\begin{figure}
\centering
\includegraphics[width=1.00\linewidth]{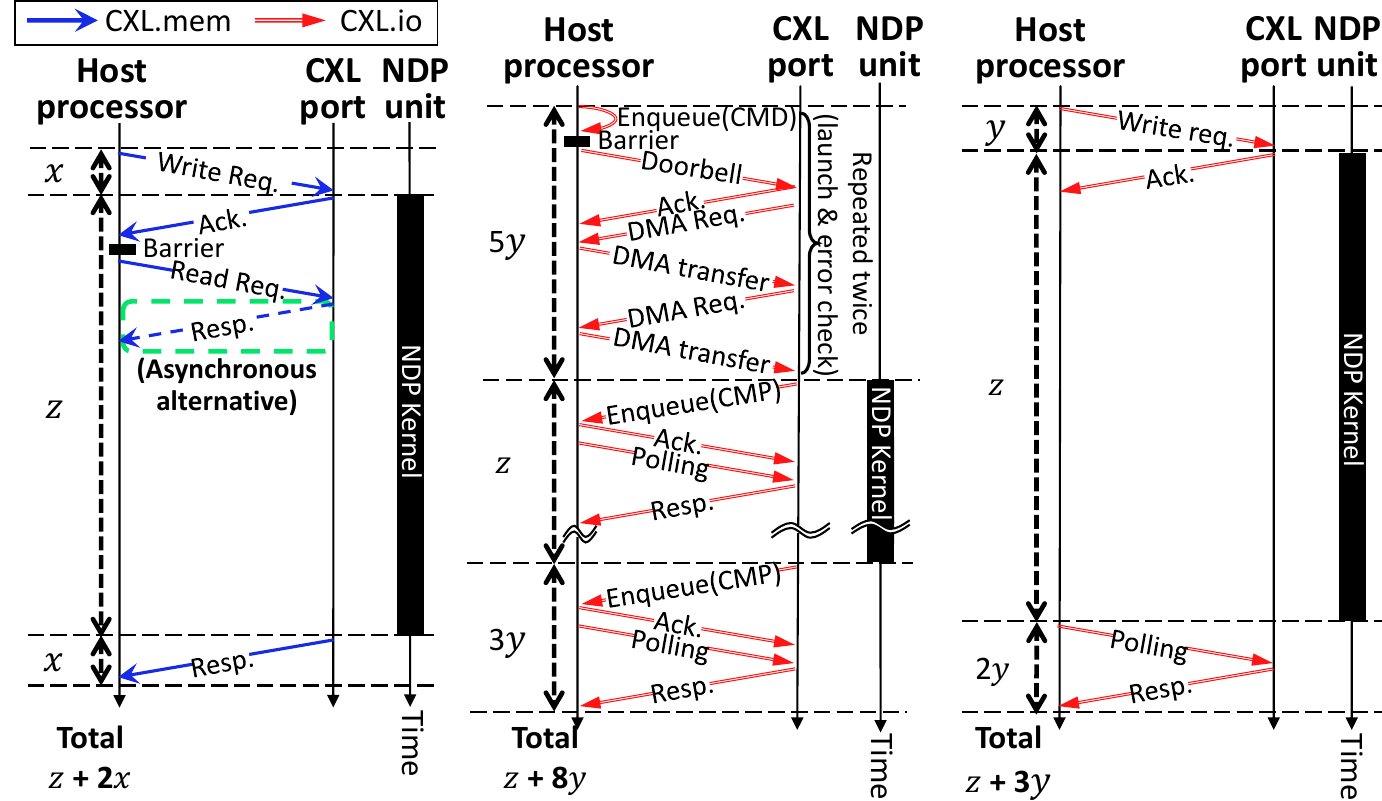}
\hbox{\hspace{0.10in}\footnotesize{ \hspace{0.05in}(a) \mmfunc{} \hspace{0.3in} (b) CXL.io (ring buffer)\hspace{0.2in} (c) CXL.io (direct)}}
\caption{Example timelines with different NDP offloading schemes.
One-way latencies of CXL.mem, CXL.io, and kernel execution are parameterized as $x$, $y$, and $z$,
respectively. Their known minimal values are $x$=$\sim$75~ns from 150~ns load-to-use latency for CXL memory~\cite{cxl_scale, pond},
$y$=$\sim$500~ns from $\sim$1~$\mu$s DMA~\cite{pcie-latency}. An example value for $z$ is
6.4~$\mu$s NDP kernel runtime from \textfw{DLRM(SLS)-B32} (\S \ref{sec:perf}). 
For \mmfunc{}, we assume a synchronous launch while also showing the arrow for an alternative asynchronous launch.
For the ring buffer, CMD and CMP refer to command and completion messages enqueued into the ring buffers,
respectively. Two pairs of CMD and CMP are needed for kernel launch and error checks~\cite{cuda_error_check}. 
While the barrier for \mmfunc{} overlaps with the kernel, the one needed for ring buffer is
in the critical path.}
\label{fig:timeline}
\end{figure}

The \mmfunc{} facilitates efficient NDP kernel launch with minimal overhead (Fig.~\ref{fig:timeline}a).
To launch a kernel, a host process can invoke the 
\mmfunc{} at offset 2$\ll$5 (Table~\ref{tab:ndp_func})
by issuing a write request with kernel launch arguments.
Note the difference between \mmfunc{} arguments for kernel launch 
function (which determines how a kernel is launched) 
and NDP \textit{kernel} arguments (which will be directly used in the NDP kernel code). 
Large kernel inputs (e.g., arrays) can be stored in a separate memory 
location in CXL memory and their pointer can be passed as an argument.
Each kernel instance is associated with a virtual memory region for an input or output data array
called \emph{$\mu$thread pool region} provided in a kernel launch call
for our  \mmuthr{} mechanism (\S\ref{sec:uthreading}).
After a kernel launch, the NDP controller sends back 
an acknowledgment packet immediately.

Afterward, the host can have a memory fence and a load instruction 
to fetch the return value for the kernel launch function at the same \mmfunc{} offset 2$\ll$5.
The difference is that this time, a read request will be sent.
Its response with the return value can be sent back differently based on the 
\textfw{Synchronicity} argument given for kernel launch:
for a synchronous launch, it will return after kernel termination, and 
for an asynchronous launch, it will return immediately (dotted arrow in Fig.~\ref{fig:timeline}a).
The asynchronous launch enables overlapping an NDP kernel with subsequent NDP kernels
launched from the same host process or other host-side computation.
Concurrent kernels can also be launched from multiple host threads,
similar to the multi-process service (MPS) of GPUs~\cite{mps}.
The host can then later use the
kernel status poll function (i.e., ndpPollKernelStatus) to check its completion. 

When the NDP unit's available resource is insufficient due to other 
kernels running, the kernel launch request will be buffered and served 
after prior kernels are completed. If the buffer is full, the kernel launch will
return an error code.

\noindent
{\bf Comparison with traditional approaches.}
With the traditional ring buffer scheme used by PCIe/CXL.io devices,
an NDP kernel launch can require multiple link round-trips to update the write pointer (i.e., doorbell),
and transfer the pointer to the command from the ring buffer and then the command itself to 
the device similar to GPU kernel launches~\cite{amd_ring_buffer,ring_buffer2} (Fig.~\ref{fig:timeline}b). 
Subsequently, to check if the launch is done without an error, 
the procedure should be repeated~\cite{cuda_error_check}.
This approach incurs high latency but allows 
concurrent execution of multiple NDP kernels.
On the other hand, a simpler approach of directly manipulating dedicated device registers through MMIO~\cite{nda}
takes a shorter latency (Fig.~\ref{fig:timeline}c) but can execute only one kernel at a time as the
registers should not be overwritten.

In contrast to these approaches, \mmfunc{} reduces the kernel launch latency by 
exploiting the faster CXL.mem protocol and avoiding kernel mode transition.
In addition to the protocol-level advantage, \mmfunc{} requires fewer round-trips
compared to the ring buffer scheme while enabling concurrent execution of multiple kernels.
As a result, for the example latencies in Fig.~\ref{fig:timeline}, 
\mmfunc{} reduces the communication overhead and end-to-end runtime by 33-75\% and 17-37\%, respectively,
compared to the traditional schemes. %

Note that while we reduce the NDP offloading overhead with CXL.mem,
we do not preclude the use of CXL.io/PCIe for NDP management
in systems where CXL.mem is not available.
For long kernels, CXL.io overhead can be well-amortized over the runtime.
The CXL-\mmndp{} can be configured to use either
the conventional CXL.io/PCIe mechanism or \mmfunc{} with CXL.mem
when the device is initialized by the OS and driver, as using both at the same
time is unnecessary.

\noindent
\textbf{API for Host-side Programming.}
For host codes, 
we propose an API for NDP that exposes
high-level functions defined in the first three columns 
of Table~\ref{tab:ndp_func}, similar to 
the APIs of existing accelerators (e.g., CUDA). 
Thus, users do not need to understand the low-level 
implementation with \mmfunc{} -- e.g., how an API call's return value is 
fetched with a subsequent CXL.mem read request or 
the offset value for each function.
Using the ndpPollKernelStatus function,
kernel status checks
and exception handling can be done using the return value.
Note that while this API demonstrates a minimal example,
it can be easily extended to include a richer set of API functions.

\subsection{Memory-mapped $\mu$threading (\mmuthr{})}
\label{sec:uthreading}

To maximize the NDP kernel's memory bandwidth utilization,
a large number of memory accesses need to be done concurrently to hide memory latency.
While out-of-order cores can perform multiple memory accesses simultaneously, it is
not suitable for cost-efficient NDP due to high control logic overhead.
Fine-grained multithreading (FGMT), especially with a large number of threads as in GPUs,
can efficiently provide high concurrency. However, GPU SM's SIMT-only execution can be 
inefficient when its threads perform redundant computation within a warp 
due to a lack of scalar operations 
(e.g., loop variable management, and address calculation)~\cite{r2d2}.

Thus, to efficiently support both scalar and SIMD operations, we adopt RISC-V ISA 
with vector extension (RVV) and modify it to support highly
concurrent FGMT-based \mmuthr{} (Table~\ref{tab:arch_comp}).
Particularly, for CPUs, the OS creates and manages threads, 
but the overhead can be tremendous for a large number of threads, especially 
if they are short-lived~\cite{thread_coarsening}, due to
$\mu$s-scale thread management delays~\cite{thread_creation1, thread_creation2}.
In addition, a CPU thread requires the entire ISA-defined register set,
so the physical register file can grow linearly with the HW thread count.
However, memory-bound workloads tend to use fewer registers than 
compute-bound workloads due to lower arithmetic intensity. 
Thus, we use GPU-style HW-managed threads without the conventional OS for CPUs
and provision the number of registers for each thread as specified by SW (i.e., compiler) 
during kernel registration (Table~\ref{tab:ndp_func}) to reduce register file
cost. 
For example, if 5 integer and 3 vector registers are needed, 
only registers \textfw{x0}-\textfw{x4} and \textfw{v0}-\textfw{v2} are used in the kernel.
We refer to this type of thread as \emph{$\mu$thread} due to its low resource usage.
Creating a $\mu$thread can be done quickly as in GPUs. 
To maximize the concurrency of $\mu$threads,
they execute in a bulk synchronous parallel manner
without any ordering guarantee, similar to GPU threads.
The $\mu$threads can also use on-chip scratchpad memory for communication. 
Thus, the NDP kernel should be written accordingly.
Despite similarities, our $\mu$threads differ from GPU threads in several ways 
besides the ISA differences and provide 
the following key \textbf{\underline{A}}dvantages \textbf{\underline{(A1-A4)}}.

\begin{table}[]
\centering
\scriptsize
\caption{Architectural differences between the CPU, GPU, and \mmndp{}.}
\setlength{\tabcolsep}{4.0pt} %
\setlength\extrarowheight{2pt}
\begin{tabular}{|l||c|c|c|}
\hline
& {\bf CPU} & {\bf GPU} & {\bf \mmndp{}}\\ \hline \hline
{\bf Thread creation} & {\bf Each thread} & Threadblock & {\bf Each $\mu$thread}\\
{\bf granularity}     &   {\bf (fine-grained)}    &   (corase-grained)  & {\bf (fine-grained)} \\ \hline
{\bf Flynn's taxonomy} & {\bf SISD + SIMD} & {SIMD (SIMT) only} & {\bf SISD + SIMD} \\ \hline
{\bf Per-thread registers} & {Fixed by ISA} & {\bf By usage} & {\bf By usage} \\ \hline
{\bf Thread creation} & {By OS} & {\bf By HW} & {\bf By HW} \\ \hline
\multirow{2}{*}{\bf Thread scheduling} & {ST/SMT/} & \multirow{2}{*}{\bf FGMT} & \multirow{2}{*}{\bf FGMT} \\
                                     & {FGMT/CGMT} &      & \\ \hline
{\bf Out-of-order exec.} & Yes or No & {\bf No} & {\bf No} \\ \hline
{\bf Scratchpad} & \multirow{2}{*}{N/A} & \multirow{2}{*}{Threadblock} & \underline{\bf All $\mu$threads run} \\ 
{\bf memory scope} &          &               & \underline{\bf on an NDP unit} \\ \hline
{\bf Thread} & \multirow{2}{*}{Process ID} & {(Threadblock ID,} & \underline{\bf Mapped $\mu$thread} \\
{\bf Identification} & & {thread ID)} & \underline{\bf pool address} \\ \hline
\end{tabular}
\label{tab:arch_comp}
\end{table}

\noindent
\textbf{\underline{(A1)}} \mmuthr{} reduces the overhead of address calculation 
in an NDP kernel compared to GPUs.
Whereas a GPU thread is identified by multidimensional threadblock and thread indices,
$\mu$threads are identified by the address it is mapped to in a $\mu$thread pool region.
The address and offset from the base of the pool region are provided 
in the first two non-zero-valued scalar registers (i.e., \textfw{x1} and \textfw{x2})
when a $\mu$thread is spawned. 
Then, the offset can also be used to access other data with different bases.
By using one of the input data arrays as a
$\mu$thread pool region (Fig.~\ref{fig:ndp_example}), 
the $\mu$thread can reduce address calculation overhead.
As memory-bound NDP kernels tend to have fewer instructions than compute-bound 
kernels, the static instruction count is reduced by 3.28-17.6\% for our evaluated workloads as a result,
compared to calculating addresses from multi-dimensional threadblock/thread dimension and indices.

In addition, we avoid the overhead of redundant address calculation
in SIMT-only GPUs by using scalar instructions and improve performance by up to 20.2\% (\S \ref{sec:scale}).
Avoiding the redundancy also reduces the register file size
requirement and the number of ALUs per NDP unit, resulting in
smaller NDP unit area. Combined with the
goal of optimizing for memory-bound workloads,
our NDP unit uses 81\% smaller register file and 
69\% less area for ALUs (\S \ref{sec:cost}). 
As a result, compared to GPU SMs, more NDP units 
can be implemented in given area to sustain
higher concurrency in memory accesses.

\begin{figure}
\centering
\includegraphics[width=0.90\linewidth]{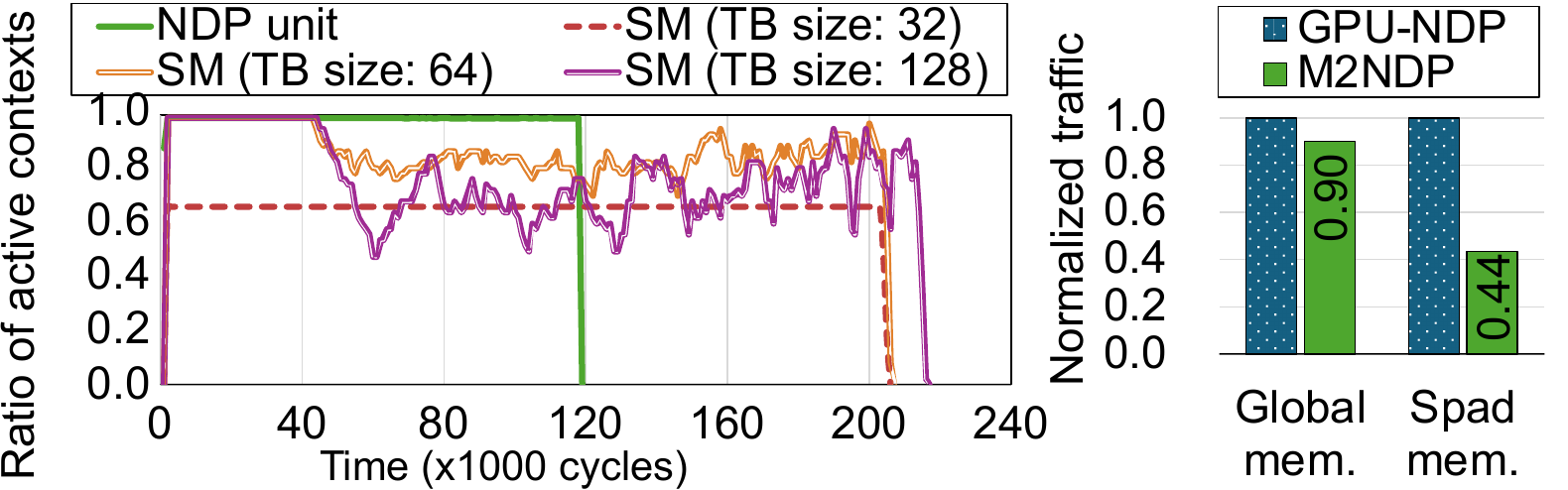}
\vspace{-.07in}
\hbox{\hspace{0.6in}\footnotesize{(a)} \hspace{1.5in} \footnotesize{(b)}\hspace{0.1in}}
\caption{(a) Ratio of active contexts (i.e., warps for GPU SMs and $\mu$threads for \mmuthr{})
executed on an SM or NDP unit over time for a main kernel of \textfw{PGRANK}~\cite{pannotia}
with configuration in \S\ref{sec:methodology}. Maximum threadblock count per SM limits
the active warp ratio for the threadblock (TB) size of 32. (b) Reduction of global and 
scratchpad memory traffic by our NDP unit for \textfw{HISTO}. For GPU-NDP, ``Iso-Area'' configuration (\S \ref{sec:methodology}) was used here.}
\label{fig:ndp_vs_gpu}
\end{figure}

\noindent
\textbf{\underline{(A2)}} Second, whereas GPU threads are created in a coarse threadblock granularity,
$\mu$threads are created in fine, individual thread granularity.
The coarse-grained thread creation can result in resource fragmentation and 
underutilization due to inter-warp divergence -- i.e., 
resource unused by finished warps of a 
threadblock will remain unused 
until the entire threadblock they belong to is finished and its resource is released for the next
threadblock~\cite{warp_diverge}.
In contrast, with \mmuthr{}, resources for a finished $\mu$thread are released immediately for
the next $\mu$thread, improving resource utilization and performance/cost 
for irregular workloads. %
For example, Fig.~\ref{fig:ndp_vs_gpu}a shows that, 
for \textfw{PGRANK}, NDP unit increases the ratio of active contexts by 50.9-15.9\% compared to
GPU SM using different threadblock sizes.
While using smaller threadblock can improve resource utilization in some cases,
it can make it more difficult to effectively use the CUDA shared memory because 
different threadblocks cannot share data through shared memory. As a result,
global memory traffic can be increased.
By removing the threadblock hierarchy, \mmuthr{} also eliminates the need for optimizing the
threadblock dimension, which can significantly affect performance~\cite{cuda_opt_gtc}.

\noindent
\textbf{\underline{(A3)}}
Moreover, the scope of the on-chip scratchpad memory in NDP unit 
is larger for $\mu$threads than in CUDA. 
Whereas CUDA shared memory is not shared across threadblocks 
even if they are executed on the same SM, all $\mu$threads 
executed on the same NDP unit can share data through 
the scratchpad memory.
As a result, our NDP unit can significantly reduce traffic
for global memory and on-chip scratchpad memory 
compared to GPUs -- e.g., 10\% and 56\%, respectively (Fig.~\ref{fig:ndp_vs_gpu}b).
Initializing the shared memory in 
each threadblock
also requires additional intra-block synchronization.
While NVIDIA's Hopper GPU~\cite{nvidia-h100} introduces distributed shared memory that allows 
different threadblocks in a threadblock cluster to share data in shared memory,
it requires that the threadblocks be scheduled in the even coarser cluster granularity
and can aggravate SM resource underutilization (Fig.~\ref{fig:ndp_vs_gpu}a).

\noindent
\textbf{\underline{(A4)}}
To achieve high utilization of the vector ALUs
while avoiding bottleneck, the size of the data associated 
with a $\mu$thread is matched with the memory access granularity of the DRAM  (e.g., 64~B for DDR5 and 32~B for LPDDR5).
In contrast, GPU tends to process a larger amount of data in a warp (e.g., 128~B per warp
with 32 threads processing FP32 data). 
As a result, for irregular workloads, there can be significant intra-warp 
divergence, lowering compute resource utilization.  
As a result, for the (irregular) graph workloads we evaluated, the proportion of active
lanes in the SIMD units was 1.39-2.27$\times$ higher in our NDP unit compared to the GPU SM.

\subsection{NDP Unit Microarchitecture}
\label{sec:ndp_unit}

\begin{figure}
\centering
\includegraphics[width=0.95\linewidth]{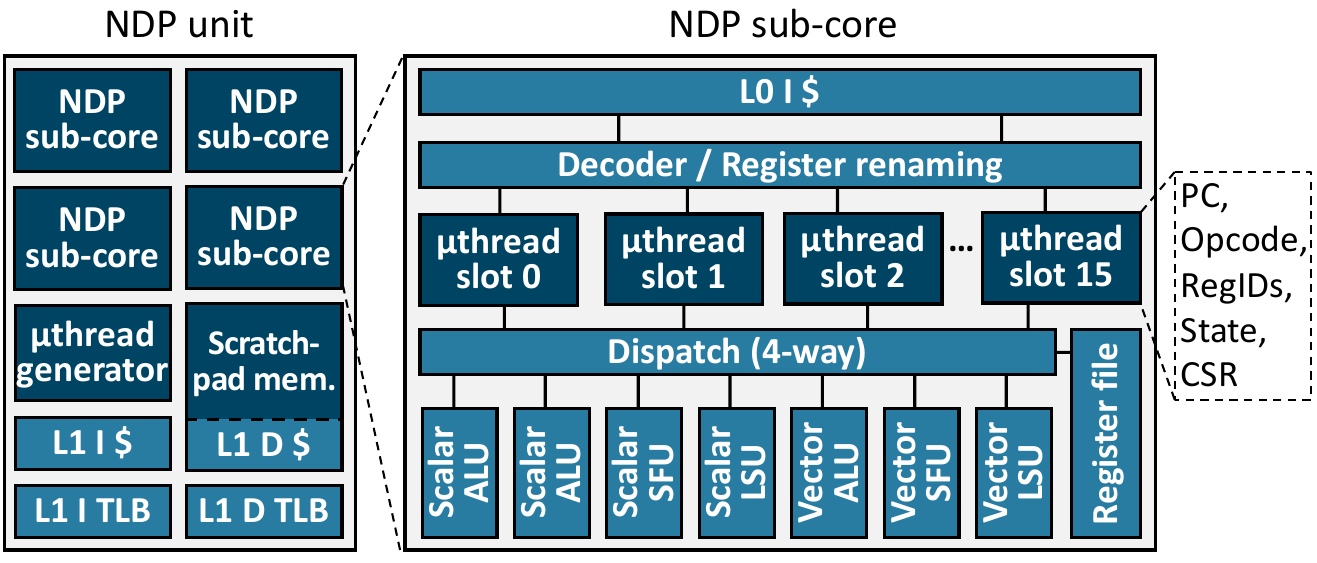}
\caption{Proposed NDP unit microarchitecture.}
\label{fig:ndp_unit}
\end{figure}

The NDP unit is designed at low cost while supporting general-purpose 
computation (Fig.~\ref{fig:ndp_unit}).
When an NDP kernel is launched, the NDP controller commands the \emph{$\mu$thread generator}
to spawn $\mu$threads by allocating \textit{$\mu$thread slots} and register file resources 
across the sub-cores of the NDP unit.
Having multiple sub-cores instead of a monolithic core simplifies the dispatch unit. 
A $\mu$thread slot consist of a PC (program counter), CSR (configure and status register) of RISC-V,
opcode and register IDs of the current instruction decoded, 
and base IDs for INT/FP/vector registers. 
The base register IDs are assigned when a $\mu$thread is created
and allocated the required registers.
Logical registers are renamed to physical registers 
simply by adding a logical ID to the base ID.

To load-balance NDP units, $\mu$threads are scheduled on NDP units
in an interleaved manner with the memory-access granularity.
Otherwise, there can be a significant load-imbalance among
NDP units for fine-grained NDP kernels 
(e.g., one NDP unit could have 64 active μthreads while the others are idle).
After a $\mu$thread is allocated a slot, its PC is initialized with 
the kernel code location to begin execution.%

A load/store unit for the scratchpad memory with atomic operations
capability~\cite{riscv-vamo} is also provided to manipulate shared data 
in an NDP unit (e.g., for reduction by multiple $\mu$threads).
Global memory atomics are done at the memory-side L2 cache 
to avoid coherence issues (\S\ref{sec:cache}).
Address translation is done using the on-chip TLBs, DRAM-TLB, and ATS (\S\ref{sec:vm}).
The NDP unit can access any memory location in CXL memories in the system 
through on-chip and off-chip interconnects.
The on-chip crossbar provides high BW
for all-to-all communication between the NDP units
and the memory controllers. 
On-chip wires and BW are 
abundant~\cite{route-packets}, 
and our crossbar is significantly
smaller than that of GPUs~\cite{a100_noc}.

Instructions from a $\mu$thread are executed serially  
while different $\mu$threads independently issue instructions with FGMT,
avoiding the overhead of complex dependency checks between instructions or data forwarding logic.
With sufficient $\mu$thread slots (e.g., 64 per NDP unit), 
the CXL memory bandwidth can be highly utilized. %
When a $\mu$thread is finished, another $\mu$thread in the $\mu$thread pool 
is spawned in the idle slot.

\subsection{Caches Hierarchy}
\label{sec:cache}

To avoid the complexity of cache coherence, we adopt the cache hierarchy of the GPU~\cite{cuda_opt_for_a100},
using write-through policy for L1 data cache of NDP units and placing the L2 cache in front 
of the memory controller (Fig.~\ref{fig:overview}).
L1 data cache's capacity is also configurable between normal L1 data cache and scratchpad memory
to meet varying requirements of different workloads.
The L2 cache supports global memory atomic operations for data from DRAM.
The NDP unit employs a small instruction cache because data-parallel, memory-bound workloads
have relatively smaller instruction footprint than compute-bound workloads.  
To prevent access to stale code, the instruction caches are flushed 
when an NDP kernel is unregistered (\S\ref{sec:management}).
However, it would be done infrequently and have negligible performance impact.

\subsection{Programming Model for NDP Kernels}
\label{sec:ndp_kernel}

\begin{figure}
\centering
\includegraphics[width=0.80\linewidth]{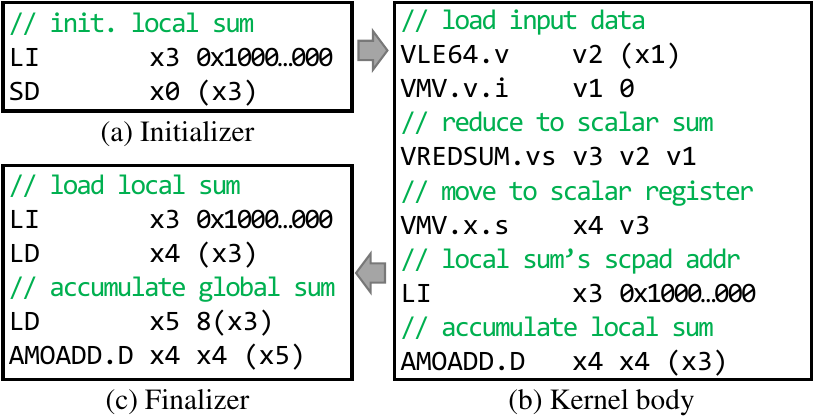}
\caption{NDP kernel example for reduction of a large data. It is assumed
that the scratchpad memory is mapped to 0x10000000 and the final result will be stored at the location 
given in scratchpad memory at 0x10000008. The 
\texttt{AMOADD} instruction performs an atomic add operation on a memory location.}
\label{fig:kernel_example}
\end{figure}

To support various use cases, an NDP kernel consists of 
an \textit{initializer}, \textit{kernel body}, and \textit{finalizer}.
The initializer (Fig.~\ref{fig:kernel_example}a) is executed only once when an NDP kernel is launched
for initialization of scratchpad memory (if needed) and any required pre-computation before
the main computation. 
For the initializer, one $\mu$thread is spawned in each $\mu$thread slot with a unique ID
in the x2 (or offset) register (\S\ref{sec:ndp_unit}).
When they are finished, the $\mu$thread generator starts spawning 
$\mu$threads from the $\mu$thread pool region to 
execute the kernel body (Fig.~\ref{fig:kernel_example}b).
There can be multiple kernel bodies such that when a kernel body is finished for all $\mu$threads,
all $\mu$threads are generated again for the next kernel body.
It can be useful for synchronization of $\mu$threads across different phases of a kernel.
After all kernel bodies finish, the finalizer (Fig.~\ref{fig:kernel_example}c)
is executed, similar to the initializer, but for post-processing 
and storing the result to DRAM if needed.

When a kernel is launched, the kernel arguments are 
placed in the on-chip scratchpad memory of each NDP 
unit 
to efficiently share them among $ \mu$threads.
The scratchpad memory is mapped to the unused region in the virtual memory 
layout~\cite{riscv-vmlayout} and can be accessed using normal loads/stores.

\mmuthr{} provides a very flexible execution environment
with few restrictions. 
Depending on the HW support, the compiler can use 
any instruction in RV64IMAFDV extension or its subset,
except for instructions that require operating system
(e.g., \textfw{ECALL}).
In addition, kernels can access any memory location in HDM, 
including that of peer CXL.mem devices, either directly or indirectly.
Thus, pointer chasing can be done for irregular workloads (e.g., graph analytics).
While host-side memory cannot be directly accessed from an
NDP kernel using CXL.mem due to the lack of support
by the protocol, 
it is possible to adopt page-fault handling support
from GPUs with PCIe and host driver/runtime.

While mapping each $\mu$thread to a memory location simplifies
the kernel code (\S \ref{sec:uthreading}), 
it is not necessary to strictly adhere
to this approach. It is even possible to map $\mu$threads to unallocated dummy
memory locations as long as they are not actually
accessed by loads/stores. In such a case, the offset in the \texttt{x2}
register can be used as a thread ID.

To generate kernel code, RISC-V compilers with 
RVV support~\cite{rvv-intrinsic} can be adapted for \mmndp{}.
For basic functionality, the compiler should assume that,
for each $\mu$thread, the $\mu$thread generator will initialize
the \textfw{x1} and \textfw{x2} registers with mapped address and offset,
respectively (\S \ref{sec:ndp_unit}).
It is also possible to adopt high-level
programming model for SIMD units in CPUs 
(e.g., Intel's ISPC~\cite{ispc, ispc-github} and DPC++/SYCL~\cite{sycl} for x86 AVX) for \mmndp{}.
Similar to CUDA, ISPC enables the SPMD programming model for vector/SIMD units 
and has been used in production and state-of-the-art graphics
frameworks~\cite{embree, embree-github, luisarender, moonray}.
Additionally, similar to how cuDNN and cuBLAS from NVIDIA are developed and 
optimized in assembly~\cite{accel_sim, fireiron}, hand-tuned \mmndp{} libraries 
can be developed to achieve high performance for common high-level operations.
Unfortunately, since RISC-V has a shorter history, 
its software ecosystem has not yet matured enough and lacks 
such open-source compilers and libraries.
We leave designing such compilers for future work.

\subsection{Virtual Memory Support}
\label{sec:vm}

Our \mmndp{} can efficiently support virtual memory.
As the host uses physical addresses for normal CXL.mem requests,
address translation is not needed in a passive CXL memory without NDP.
However, with NDP, virtual addresses are used for the $\mu$thread pool region and
load/store instructions. 
Our NDP unit employs on-chip TLBs, but
it may be insufficient for kernels that 
process large data in CXL memory, and the ATS (\S \ref{sec:cxl_background})
can also incur high latency.
Thus, we adopt DRAM-TLB~\cite{pom-tlb, ducati} to cost-effectively improve 
the TLB reach of NDP units and minimize the miss penalty of on-chip TLBs.

Each DRAM-TLB entry uses 16 bytes to store the ASID, tag, physical page number, 
and other attributes (e.g., permission bits).
The location of a DRAM-TLB entry is computed based on the hash of the virtual page number and
ASID as well as base address per CXL memory, ensuring 
that all NDP units within 
the same CXL memory can share them.

The DRAM-TLB has low overhead since 
even with 4~KB pages, the  
DRAM-TLB entry has only 16~B/4~KB=0.4\% overhead, and for 2~MB pages, the overhead is negligible.
If the DRAM-TLB region is sufficiently allocated 
for the given capacity of CXL memory, 
there will be few DRAM-TLB misses 
with the hashed location calculation, after DRAM-TLB warms up.

The on-chip and DRAM-TLBs of CXL-\mmndp{} can also keep translations for addresses in other CXL memories if they exist.
A TLB shootdown needs to be done for all CXL-\mmndp{}s if a page's mapping changes, 
but it can rarely occur for in-memory data we assume (i.e., no swapping to disks).

\subsection{Scaling with Multiple CXL-\mmndp{}s}

Using direct P2P accesses between CXL devices through a CXL switch (\S\ref{sec:cxl_background}),
NDP kernels can access data from other CXL-\mmndp{}s to process huge data.
However, the CXL interface bandwidth can become a bottleneck for frequent P2P accesses,
so localizing data across multiple CXL memories needs to be done carefully.
Because different workloads exhibit varying memory access patterns, data partitioning
schemes are typically specialized for target workloads~\cite{megatron-lm}. %
For best performance, current multi-GPU systems also require the user-level 
SW to partition the data across GPUs and launch separate kernels.
Thus, we similarly assume that the data are placed by SW across CXL memories
and an NDP kernel is launched in each CXL-\mmndp{} for multi-device scaling,
and leave the exploration of automatic scaling for future work.
However, the data localization does not have to be perfect since NDP units can directly
access other CXL memories for reads and atomic operations similar to GPUs. %

\begin{figure}
\centering
\includegraphics[width=1.0\linewidth]{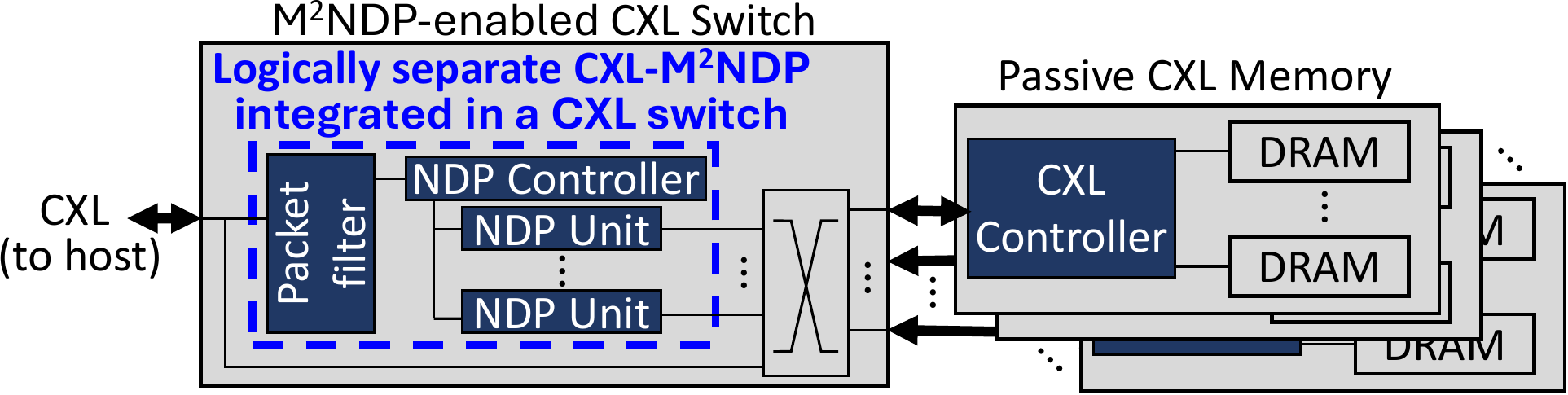}
\caption{CXL switch with integrated \mmndp{} logic that can process data from passive CXL memories.}
\label{fig:switch-ndp}
\end{figure}
\subsection{Scaling CXL Memory Capacity Independently of NDP Units with an \mmndp{}-enabled CXL Switch }
\label{sec:passive_cxl}
Using multiple CXL-\mmndp{}s increases total NDP throughput proportionally with 
the total CXL memory capacity, which can be
desirable in many cases. However, some workloads may have low throughput/capacity ratios
and need to increase capacity independently of NDP throughput.
For such scenarios, CXL-\mmndp{} can be integrated in a CXL switch
to perform NDP with data from different peer, (third-party) passive CXL memories
(Fig.~\ref{fig:switch-ndp}). For \mmfunc{} region (\S \ref{sec:overview}),
a small SRAM within the switch can be used. 
To avoid coherence issues with host, it is desirable to use it for workloads
that do not need concurrent shared data manipulation
between the host and NDP (e.g., serving ML models).

\begin{table}
\caption{Simulator configuration. When multiple values are given, the default is indicated in boldface.}
\scriptsize
    \centering
    \renewcommand{\arraystretch}{1.00}
\setlength{\tabcolsep}{2.5pt} %
    \begin{tabular}{l|l}
    \hline
    \multicolumn{2}{c}{ \bf GPU} \\ 
    \hline
    {\bf Parameter} & {\bf Value}   \\
    \hline
    SM count and freq. & 82 SMs @ 1695 MHz \\
    \hline
    \multirow{3}{*}{SM organization} & Max. 32 threadblocks, Max. 1536 threads, 256~KB reg. file,\\
      & 4 SP units, 4 DP units, 4 SFU units, 4 INT units, \\
      & 4 INT units, 4 TC (tensor core) units \\
    \hline
    L1 D-cache & 128 KB per SM, 128 B line, 32 B sector @ 1695 MHz \\
    \hline
    L2 cache & 6 MB per GPU, 128 B line, 32 B sector @ 1695 MHz\\
    \hline
    NoC & 82x48 crossbar (32B flit)\\
    \hline
    DRAM (HBM2) & 32 channels, 4 bankgroups/channel,\\
    organization and & 4 banks/bankgroup, tRC=48, tRCDR=14, tRCDR=10, \\
    timing param. in clk & tCL=14, tRP=15, tCCDs=1, tCCDl=2, Freq=1000~MHz \\ %
    \hline 
    \hline
    \multicolumn{2}{c}{ \bf CPU}  \\ 
    \hline
    {\bf Parameter} & {\bf Value}   \\
    \hline
    Cores & 64 OoO cores @ 3.2 GHz \\
    \hline
    \multirow{3}{*}{Caches} & 64~KB L1 (8-way, 4-cycle; 64~B line, LRU), \\
     & 1~MB L2 (8-way, 12-cycle, 64~B line, LRU), \\
     & 96~MB L3 (16-way, 74-cycle, 64~B line, LRU)  \\
    \hline
    DRAM (timing &  DDR5-6400 with 409.6~GB/s (8 channels)\\
    parameters in clk) & tRC=149, tRCD=46, tCL=46, tRP=46\\
    \hline
    \hline
    \multicolumn{2}{c}{\bf CXL Memory Expander} \\ 
    \hline
    {\bf Parameter} & {\bf Value}   \\
    \hline
    \multirow{2}{*}{CXL} & 64~GB/s (in each dir.) from CXL 3.0 (PCIe 6.0) x8, 256~B flit\\
    &Load-to-use latency: {\bf 150~ns}, 300~ns, 600~ns\\
    \hline
        NoC & Four 32x32 crossbars (32B flit)\\
    \hline
    Memory-side  & 4~MB (128~KB per memory channel,\\
    L2 cache & 16-way, 7-cycle, 128~B line, 32~B sector, LRU)\\
    
    \hline
    DRAM (timing & 32-channel LPDDR5 with 409.6~GB/s and 
    256~GB capacity \\
    parameters in clk) & per device~\cite{cxl-pnm}, tRC=48, tRCD=15, tCL=20, tRP=15\\
    \hline
    \hline
    \multicolumn{2}{c}{\bf NDP in CXL Memory} \\ 
    \hline
    {\bf Type} & {\bf Configuration}   \\
    \hline
    \multirow{8}{*}{\shortstack[l]{\mmndp{} \\ (SC: sub-core)}} & 32 NDP units @ 2~GHz, 4 SCs per NDP unit,\\
     &48~KB register file, 512~B L0 I-cache per SC, \\
    & 2~KB L1 I-cache, 128~KB scratchpad/L1D cache \\
                    & (16-way, 4-cycle, 128~B line, 32~B sector), \\
                    &256-entry I-TLB, 256-entry D-TLB (8-way),\\
     & Scalar units: 2 ALUs, 1 SFU, and 1 LSU per SC,\\
     & 256-bit vector units: 1 vALU, 1 vSFU, and 1 vLSU per SC,\\
     & 16 $\mu$thread slots per SC, Max. concurrent kernels: 48 \\
    \hline
     & Iso-FLOPS(8SMs), $4\times$FLOPS(32SMs), 16$\times$FLOPS(128SMs),  \\
   GPU-NDP & Iso-Area(16.2SMs) @2~GHz,\\
    & SM organization: same as the above GPU SM without TC\\
    \hline
    \end{tabular}
    \label{tab:config}
\end{table}

\section{Evaluation}
\subsection{Methodology}
\label{sec:methodology}

We faithfully modeled the functional and timing aspects of CXL-\mmndp{} with an 
in-house cycle-level simulator based on Ramulator~\cite{ramulator}.
Baseline CPU and GPU with passive CXL memory are modeled using modified 
ZSim~\cite{zsim} and Accel-Sim~\cite{accel_sim};
while CPUs are typically used as hosts, 
for data-parallel GPU workloads, we assume GPU as the host processor because 
GPUs integrated with CPU cores can function as a host~\cite{mi300-cxl}. 
Table~\ref{tab:config} gives the simulator configurations.
The baseline GPU has similar throughput and memory bandwidth to the NVIDIA GeForce RTX 3090 GPU~\cite{nvidia-ga102}.
In addition, we provide comparisons with high-end CPU~\cite{epyc-75F3} and GPU cores~\cite{nvidia-ga102}
used for NDP within CXL memory, referred to as \textfw{CPU-NDP} and \textfw{GPU-NDP}, respectively. They represent prior approaches for general-purpose NDP. 

For \textfw{CPU-NDP} evaluation for \textfw{OLAP} workload,
we measure the performance on a real dual-socket system with 
high-end AMD EPYC 75F3 CPUs (2.3~GHz)~\cite{epyc-75F3} that has the same total memory bandwidth 
as the CXL memory that we model (i.e., 409.6~GB/s).
The evaluation was done using multiple copies of Apache Arrow processes and
memory allocation was done locally to avoid the NUMA effect.
We used 32 CPU cores in total (i.e., 16 cores per socket) to match the 32 NDP units
we assume for \mmndp{}. Note that \mmndp{} has substantially lower cost than this CPU 
with OoO pipeline and large caches.

The \textfw{\gpundpEq} uses eight Ampere GA102 SMs that provide equivalent peak FLOPS
as the 32 NDP units in CXL-\mmndp{}. 
\textfw{\gpundpQuad} and \textfw{\gpundpHexa} are also evaluated to show the impact of 
4x and 16x higher SM counts (i.e., 32 and 128 SMs). 
For \textfw{\gpundparea}, we estimate the GPU SM's area using the same methodology 
as NDP unit (\S \ref{sec:cost}) to obtain GPU-NDP with 16.2 SMs that has similar area as \mmndp{}.
We used 16 SMs and increased the SM's frequency to account for the remaining 0.2 SMs.
We also model prior work on GPU-like general-purpose NDP~\cite{ndp_sc} which requires the host to translate and generate all memory addresses for NDP
(\textfw{\scndp{}}).
All configurations except for \mmndp{} use CXL.io for kernel launch.
The direct MMIO scheme (Fig.~\ref{fig:timeline}c), which uses dedicated device
registers with a 1.5~$\mu$s latency overhead, is the default for CXL.io and is 
indicated with the \textfw{DR} suffix.
The \textfw{RB} suffix indicates the ring buffer 
scheme with a 4~$\mu$s latency overhead (Fig.~\ref{fig:timeline}b).
The \mmndp{} configuration uses CXL.mem-based \mmfunc{} for kernel launches
with CXL.mem latency according to Table~\ref{tab:config}.
All results include the communication overhead through CXL.io/CXL.mem-based mechanisms.
Unless otherwise mentioned, we evaluate performance for running a single instance of each workload at a time,
but for throughput measurements with \textfw{DRLM} and \textfw{KVStore} (\S \ref{sec:workloads}), multiple kernel instances are
executed concurrently.

In the CXL memory, we assume fine-grained 256~B-granularity hashed interleaving 
across memory channels~\cite{ipoly}.
For multiple CXL memories, we assume each page (2~MB)
is mapped to a single CXL memory as in current NUMA or multi-GPU systems~\cite{unified_memory}.
We assume locality-aware page placement by the user
for regular workloads in our evaluation. 
We assume the DRAM-TLB is warmed up for the CXL memory-resident data.

The CPU energy is modeled with McPAT~\cite{mcpat} and for GPU and NDP units, we use AccelWattch~\cite{accelwattch}, CACTI 6.5~\cite{cacti, cacti6.5} for SRAM,  
DSENT~\cite{dsent} for NoC,
and 8~pJ/bit CXL link energy~\cite{gtc2020}.
During NDP, the idle host's energy is included.

\subsection{Workloads}
\label{sec:workloads}

We focus on important workloads, including in-memory OLAP, NoSQL, 
graph analytics, and deep learning 
that exhibit large memory footprint and little cache locality (Table~\ref{tab:workload}).  
We assume that the host does not have dirty cachelines 
for the NDP kernel data by default, but show dirty host cache's impact
in \S \ref{sec:scale}.
Since the compiler for \mmndp{} is not available yet, the kernels were implemented with assembly.

\noindent

\begin{table}[]
\caption{Workloads used for evaluation (B: Baseline)}
\centering
\scriptsize
\setlength{\tabcolsep}{4.4pt} %
\begin{tabular}{c|c|l|l}
\hline
\textbf{B} & \textbf{Workload}  & \textbf{Input problem} & \textbf{Data in CXL mem.}  \\ \hline \hline
\multirow{4}{.7em}{C P U} & \multirow{2}{*}{\texttt{OLAP}~\cite{tpc-h, ssb}} & TPC-H (Q6, Q14), & Arrow columnar \\
                   & & SSB (Q1.1, Q1.2, Q1.3) & format table\\ \cline{2-4}
                   & \multirow{2}{*}{\texttt{KVStore}~\cite{redis}} & 24B key, 64B value, & Hash table with  \\
                   & & 10M KV items & key-value pairs\\ \hline
\multirow{9}{.7em}{G P U} & \multirow{2}{*}{\texttt{HISTO}~\cite{cuda_samples}} & 16M INT32 elem., & \multirow{2}{*}{Input array} \\
                   & & 256 or 4096 bins & \\ \cline{2-4}
                   & \multirow{2}{*}{\shortstack[c]{\texttt{SPMV}~\cite{prim}}} & \multirow{2}{*}{28924 nodes, 1036208 edges} & Sparse CSR matrix, \\
                   & & & dense vector \\ \cline{2-4}
                   & \makecell{\texttt{PGRANK}~\cite{pannotia}} & 299067 nodes, 1955352 edges & CSR format graph \\ \cline{2-4}
                   & \makecell{\texttt{SSSP}~\cite{pannotia}} & 264346 nodes, 733846 edges & CSR format graph \\ \cline{2-4}
                   & \makecell{\texttt{DLRM}~\cite{dlrm}} & 1M 256-dim. vectors, 256 req.& Embedding table \\ \cline{2-4}
                   & \multirow{2}{*}{\shortstack[c]{\texttt{OPT}~\cite{transformer_opt}}} & OPT-30B, OPT-2.7B, Generation & Model weight, \\
                   &  & phase with context length of 1024 & activation \\ \hline
\end{tabular}
\label{tab:workload}
\end{table}

\noindent
{\bf In-memory OLAP.} 
Filtering operations are commonly used in OLAP, but executing them from the host 
processor can cause a bottleneck in the CXL link.
Thus, using NDP, we offload the memory-intensive \textfw{Evaluate} phase of the filtering operation, 
which sweeps column data to check the filtering condition and generates 
a boolean mask in the CXL memory. 
For baseline, we use Polars~\cite{polars}, a high-performance columnar in-memory query engine
based on Apache Arrow~\cite{apache_arrow}.
A subsequent \textfw{Filter} phase (creating a resulting filtered column) and
other parts of query execution (e.g., query planning) can be efficiently executed
on the host due to small memory footprints. 
We select queries from TPC-H~\cite{tpc-h} and SSB (Star Schema Benchmark)~\cite{ssb} that
spend non-negligible time on filtering operations. 
To filter multiple columns, multiple NDP kernels are launched.
The address range of the column data is used as the $\mu$thread pool region. 

\noindent
{\bf KVStore.}
For large KVStores, the CXL memory can store hash tables and 
key-value pairs~\cite{cxl_uiuc, redis, memcached}.
Serving a KVStore request in such systems can require memory access through the CXL link for
hash table lookup, key comparison, and linked list traversal (for hash collisions). 
Thus, the tail response latency can be increased for the baseline,
but NDP can minimize data movement over CXL by offloading hash table lookup, 
reducing tail latency. 
We model a simplified Redis and offload GET/SET 
operations with NDP after compute-intensive hash function on the host. 
Request traces are obtained using YCSB~\cite{ycsb} and have 
10K requests for varying GET:SET ratios (G50:S50 for \textfw{KVS\_A} and G95:S5 for \textfw{KVS\_B}).

\noindent
{\bf Graph analytics.}
Large graph analytics require high memory capacity~\cite{graph500} and can exploit CXL memory.
As for the $\mu$thread pool region, we use the address range of the row pointers 
from the graph's CSR format.  %
Each NDP kernel corresponds to a kernel in CUDA benchmarks~\cite{prim, rodinia, parboil}.

\begin{figure*}
\centering
\includegraphics[width=0.332\linewidth]{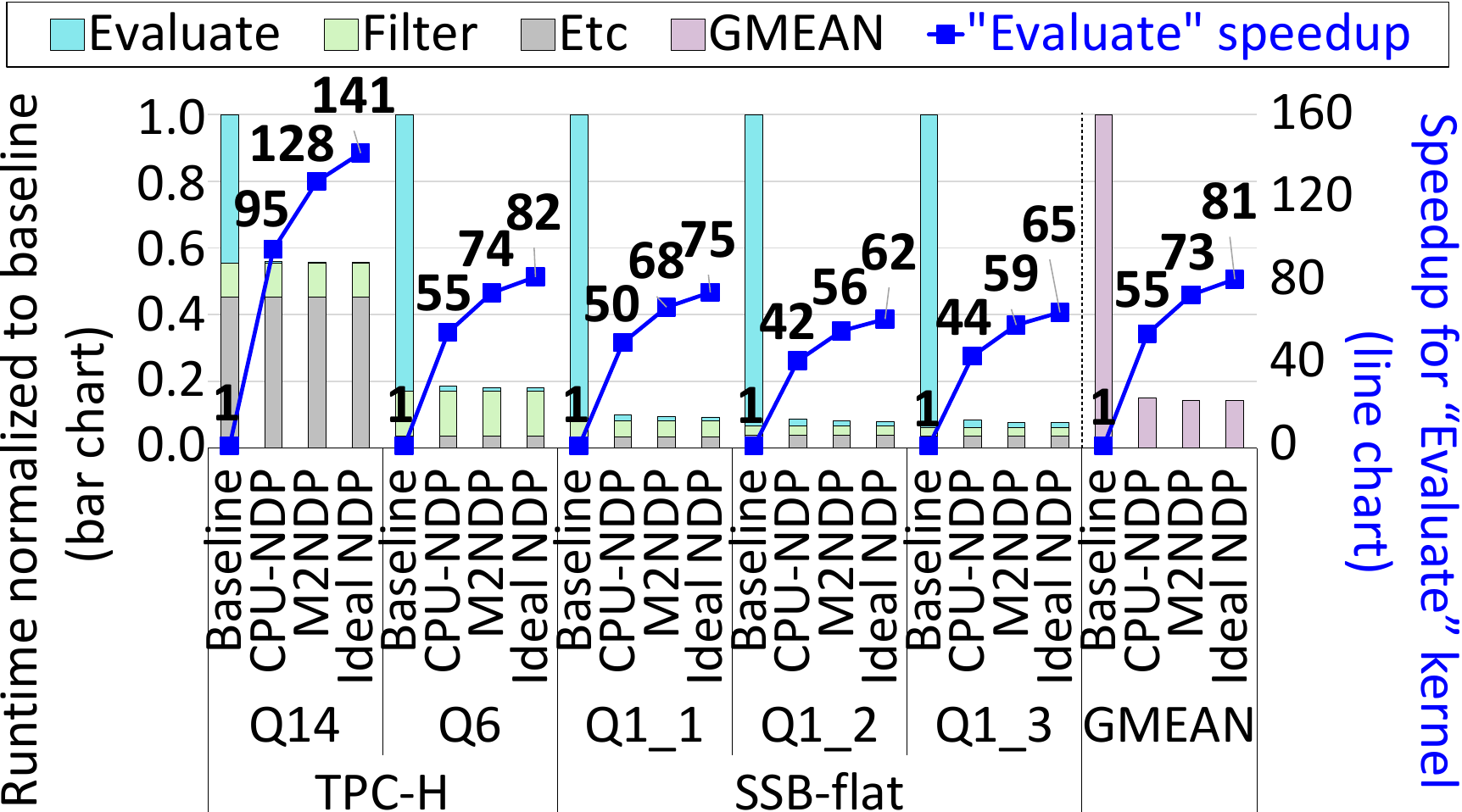}
\hspace{0.08in}
\includegraphics[width=0.128\linewidth]{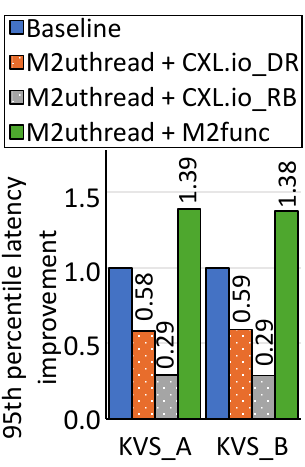}
\hspace{0.07in}
\includegraphics[width=0.481\linewidth]{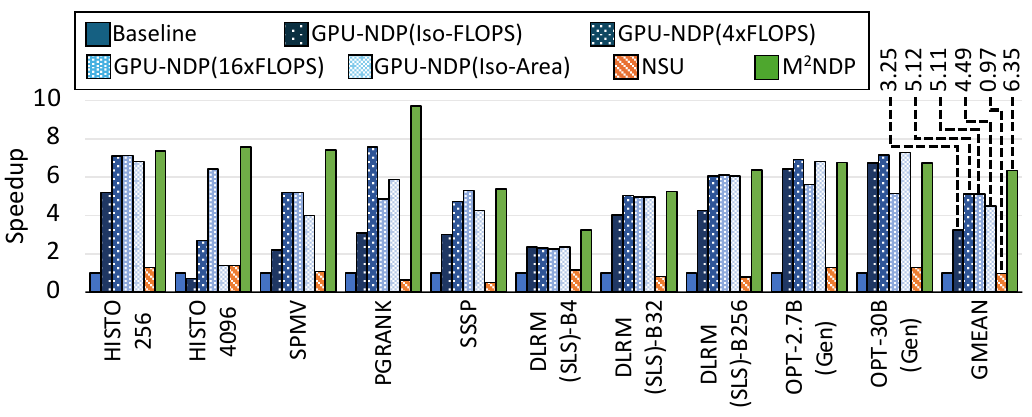}
\hbox{\hspace{0.7in}\footnotesize{(a)} \hspace{1.7in} \footnotesize{(b)}\hspace{2.1in} \footnotesize{(c)}\hspace{1.0in}}
 \vspace{-.10in}
\caption{Speedup of different NDP approaches over the baseline CPU/GPU with passive CXL memory 
for (a) OLAP, (b) KVStore, and (c) GPU workloads.}
\label{fig:perf}
\end{figure*}

\noindent
{\bf DLRM.}
Recommendation models can account for over 79\% of inference cycles in datacenters~\cite{dlrm_hpca20}.
The CXL memory can be used to cost-effectively 
store their TB-scale embedding tables~\cite{recssd}.
However, the CXL link can be a bottleneck when the host accesses the embedding tables 
for the Sparse Length Sum (SLS) operations, which can account for
up to 80\% of runtime~\cite{dlrm}.
Thus, we offload it with NDP, using the output vector of SLS as 
$\mu$thread pool region. 
We use Criteo Dataset~\cite{criteo_dataset} for input with 80 embedding lookup operations per request~\cite{recnmp}
and use batch sizes of 4, 32, and 128.

\noindent
{\bf LLM inference.}
Generative LLMs require large memory capacity from weight matrices and the key-value cache that
grows linearly with the context length during the generation phase~\cite{splitwise}. 
In addition, as GPUs are not efficiently utilized during the long generation phase~\cite{jin2023s3}, 
recent work proposed running this phase separately on GPUs with lower cost~\cite{splitwise}.
Thus, we evaluate NDP for a token generation with Meta's OPT-2.7B and OPT-30B models~\cite{meta_opt}
assuming a batch size of 1 and KV cache of 1024 tokens.
For the GPU baseline, we use the highly optimized inference kernels 
from vLLM~\cite{paged_attention}, and NDP kernels are implemented similarly.

\subsection{Performance}
\label{sec:perf}

\noindent
{\bf CPU workloads.}
Compared to the CPU baseline, for the \textfw{Evaluate} phase of \textfw{OLAP}, \mmndp{} achieved significant speedups of up to 128$\times$
(73.4$\times$ on average) with a high 90.7\% CXL memory's internal DRAM BW utilization on average (Fig.~\ref{fig:perf}a).
\mmndp{} even approached within 10.3\% of the performance of the \textfw{Ideal NDP} that uses 100\% of DRAM BW.
Our NDP units also outperformed the \textfw{CPU-NDP} with 32 high-end CPU cores with large caches~\cite{epyc-75F3} by 34.2\% on average.
For KVStore, compared to the baseline, \mmuthr{} with CXL.io-based offloading resulted in significant 1.70-3.46$\times$ increase in end-to-end P95 latency due to $\mu$s-scale
CXL.io latencies, which was significantly longer than 0.77$\mu$s P95 kernel runtime (Fig.~\ref{fig:perf}b).
In contrast, \mmfunc{} effectively improved the end-to-end P95 latency through NDP  by 38.2\% and 4.79$\times$ on average over baseline and CXL.io(RB), respectively.

\noindent
{\bf GPU workloads.}
\mmndp{} achieved significant speedups of up to 9.71$\times$ (6.35$\times$ on average) 
compared to the baseline GPU by avoiding the CXL link BW bottleneck (Fig.~\ref{fig:perf}c).
By better
utilizing resources and reducing host communication overhead, our 32 NDP units (\textfw{\mmndp{}}) even outperformed 128-SM \textfw{\gpundpHexa{}} by 24\%. 
In addition, \mmndp{} significantly outperformed \textfw{\gpundparea{}}
by up to 5.48$\times$ and 1.41$\times$ on average.
For \textfw{hist4096}, the limited threadblock-wide scope of GPU's shared memory
resulted in high global and shared memory traffic and frequent intra-block synchronization. By addressing them, \mmndp{} 
outperformed \textfw{\gpundparea{}} by 5.48$\times$.
The relative performance for graph workloads depended on the characteristics of the graph data/algorithm.
While our NDP unit used four separate 256-bit SIMD units, a GPU SM issued instructions in 32-thread warp
granularity, which was equivalent to 1024-bit SIMD width for 32-bit data. 
Thus, for the irregular graph workloads, the SMs suffered more from memory divergence depending
on the graph structure. 
For DLRM with small batch size of 4 that has short kernel runtime, \mmndp{} achieved a 37.8\% speedup over \textfw{\gpundparea{}}
by reducing kernel launch overhead.
For large-batch \textfw{DLRM} and \textfw{OPT}s,
both \textfw{\gpundparea{}} and \mmndp{} similarly 
outperformed the baseline by avoiding the CXL link BW bottleneck.
\textfw{\gpundpHexa{}} did not perform well for them
due to reduced DRAM row buffer locality caused by excessive
traffic from too many SMs. 
\textfw{\scndp{}} performed worse than the baseline on average, because the CXL link became the bottleneck
due to all addresses translated and sent from the host. 
In contrast, \mmndp{} did not have such a bottleneck,  outperforming \textfw{\scndp{}} by 6.52$\times$.

\begin{figure}
\centering
\includegraphics[width=0.39\linewidth]{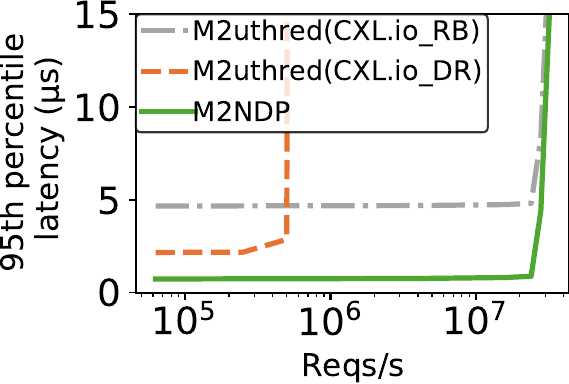}
\includegraphics[width=0.59\linewidth]{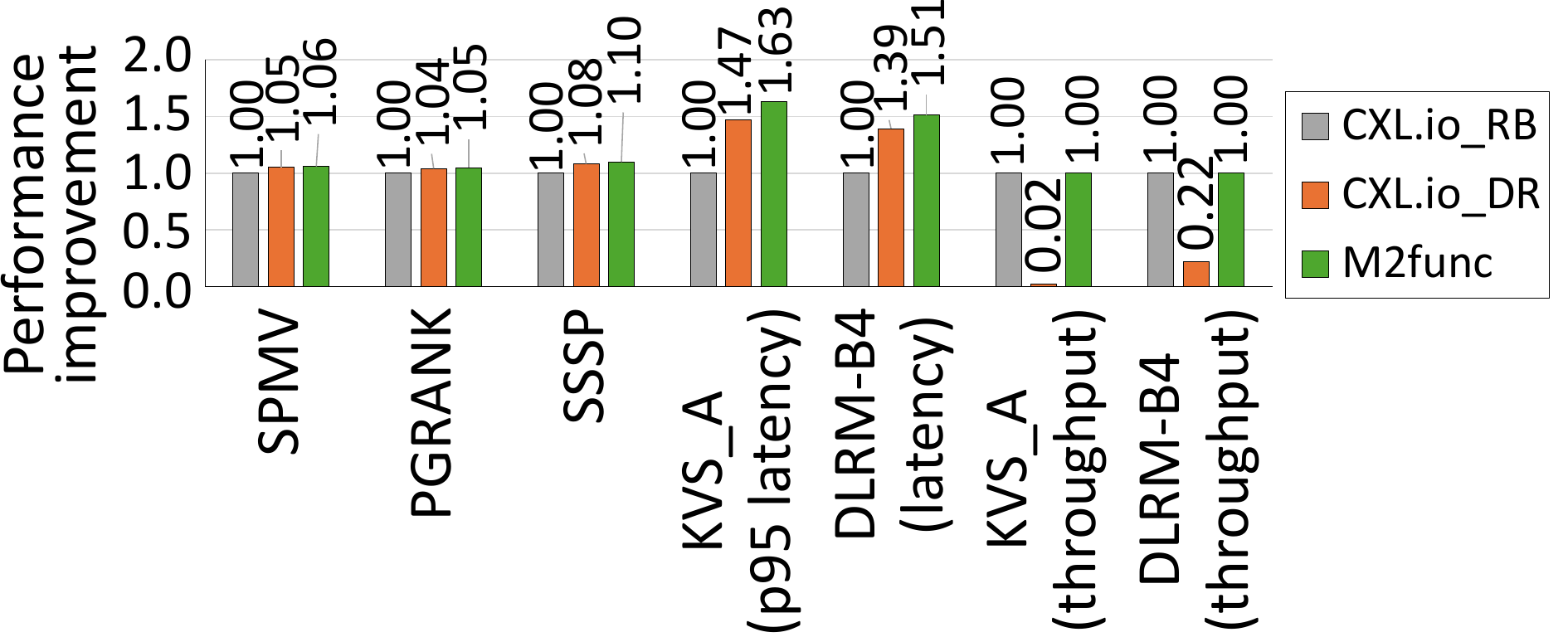}
\hbox{\hspace{0.3in}\footnotesize{(a)} \hspace{1.5in} \footnotesize{(b)}\hspace{0.3in}}
\vspace{-.09in}
\caption{(a) P95 latency-throughput curves of \textfw{KVS\_A} with latency assumptions in \S \ref{sec:methodology}.
(b) Impact of \mmfunc{} when CXL.io and CXL.mem have the same 600~ns latency.}
\label{fig:m2func_impact}
\end{figure}

\noindent
{\bf Impact of \mmfunc{}.}
By using low-overhead \mmfunc{} for host communication, \mmndp{} achieved an additional speedup of up to 2.41$\times$ (23.8\% overall) for GPU workloads over \mmuthr{} with CXL.io(RB). It was particularly effective for fine-grained NDP kernels.
In addition, compared to CXL.io(DR) that  
cannot support concurrent NDP kernels (\S \ref{sec:ndp_launch}), \mmfunc{} improved throughput by 47.3$\times$ for KVStore (Fig.~\ref{fig:m2func_impact}a).
Even if CXL.mem was assumed to have the same latency as CXL.io, 
\mmfunc{} improved latency by up to 63\% (12.1\% overall)  
over CXL.io(RB) by reducing CXL round-trips (Fig.~\ref{fig:m2func_impact}b), and increased throughput 
by 47.3$\times$ and 4.58$\times$ for \textfw{KVS\_A} and \textfw{DLRM-B4}, respectively, over CXL.io(DR), by supporting
concurrent NDP kernels.

\subsection{Scalability and Sensitivity Study}
\label{sec:scale}

\begin{figure}
\centering
\includegraphics[width=1\linewidth]{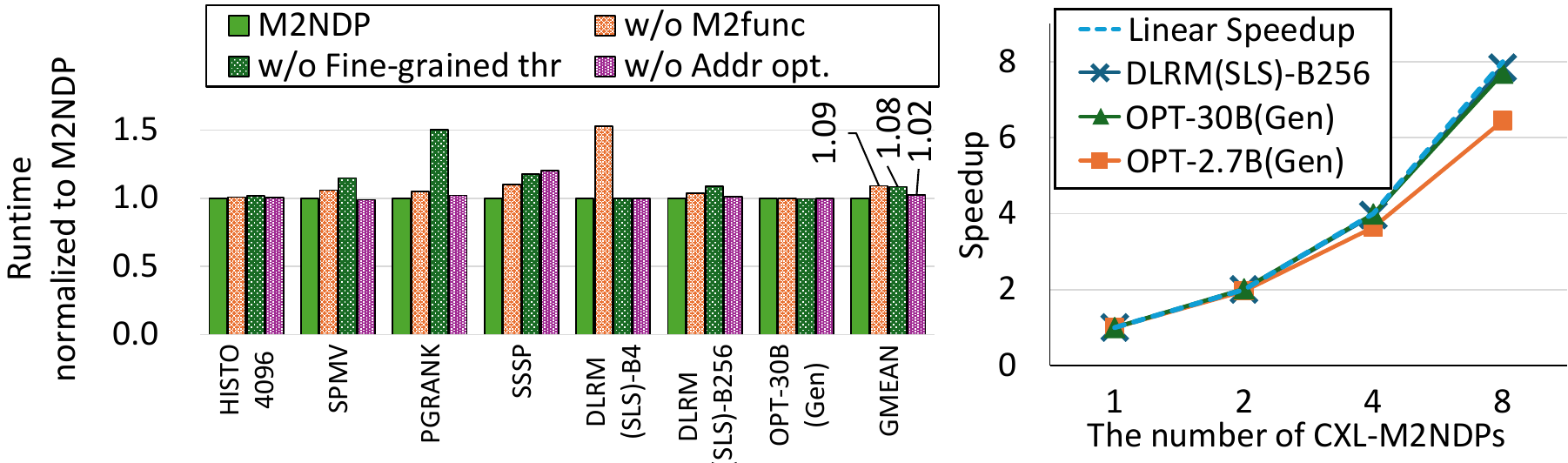}
 \hbox{\hspace{0.9in}\footnotesize{(a)} \hspace{1.5in} \footnotesize{(b)}\hspace{0.1in}}
 \vspace{-.10in}
\caption{(a) Ablation study. (b) Scalability of CXL-\mmndp{}.
}
\label{fig:ablation_scalability}
\end{figure}

\noindent
{\bf Ablation study.}
To evaluate the benefit of different components of \mmndp{}, we compare its performance with
alternative design choices (Fig.~\ref{fig:ablation_scalability}a).
Disabling \mmfunc{} and using CXL.io(RB) increased runtime by up to 141\%.
In addition, using coarse-grained $\mu$thread scheduling that spawns all 16 $\mu$threads in a sub-core
at a time increased runtime by up to 50.6\%. 
Avoiding the redundant address calculation of SIMT-only GPU by using scalar units 
had an impact of up to 20.2\%.

\begin{figure}
\centering
\includegraphics[width=1\linewidth]{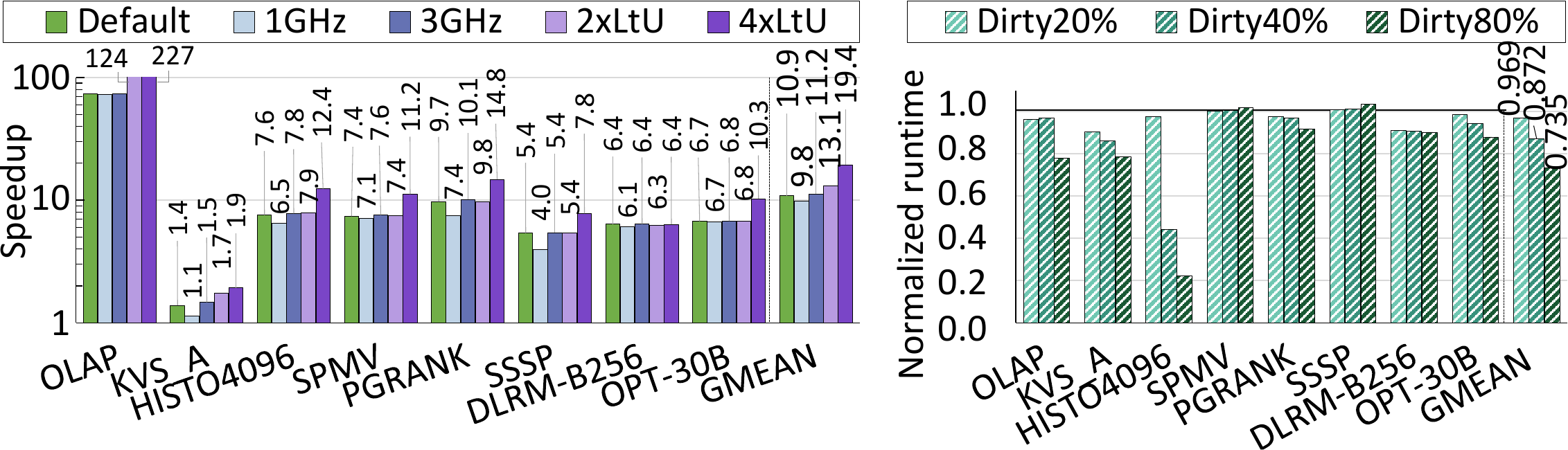}
 \hbox{\hspace{0.4in}\footnotesize{(a)} \hspace{1.5in} \footnotesize{(b)}\hspace{0.1in}}
 \vspace{-.075in}
\caption{(a) Speedup over the baselines by CXL-\mmndp{} across different NDP unit frequencies
and Load-to-Use (LtU) CXL memory latencies (\textfw{2xLtU}=300~ns, \textfw{4xLtU}=600~ns).
(b) Normalized runtime with dirty cacheline ratios over clean host cache.
\textfw{OLAP(Eval)} is the average from all queries' \textfw{Evaluate} part. 
For \textfw{KVStore}\textfw{(KVS\_A)}, we show p95 latency improvement.}
\label{fig:sensitivity}
\end{figure}

\noindent
{\bf Scalability.} 
To evaluate the scalability of \mmndp{} for \textfw{OPT} and \textfw{DLRM},
we partition the weight matrix or embedding table across different CXL-\mmndp{}s
using model parallelism~\cite{megatron-lm}.
As shown in Fig.~\ref{fig:ablation_scalability}b, we achieved near-linear speedups of 
7.84$\times$ (7.69$\times$) for \textfw{DLRM} (\textfw{OPT-30B}) 
with eight CXL-\mmndp{}s.
\textfw{OPT-2.7B} scaled less well with 6.45$\times$ speedup for 8 devices
because all-reduce took a longer portion for smaller models.

\noindent
{\bf Sensitivity study.} 
Reducing the frequency of NDP units from 2~GHz to 1~GHz degraded performance by 10.0\% overall (Fig.~\ref{fig:sensitivity}a), but using 3~GHz
improved performance by only 2.5\%
due to the memory BW bottleneck.
When load-to-use latency for CXL memory (from the host) was increased by 2-4$\times$ 
(\textfw{2xLtU} and \textfw{4xLtU}), 
the speedups by \mmndp{} further increased to 13.1$\times$ and 19.4$\times$ on average,
respectively, because the baseline suffered even more from the longer latency whereas
\mmndp{} kernels do not use the CXL link during execution and are unaffected by its latency.

In addition, when the host cache had a significant amount of 
dirty cachelines for 20-80\% of the NDP kernel's data, 
\mmndp{} was affected by only by 3.1-26.5\% overall (Fig.~\ref{fig:sensitivity}b).
Note that these scenarios are very unlikely as the host is not supposed 
to update the kernel data (e.g., LLM weights and DLRM embedding table during inference),
and the kernel data are significantly larger than the host's cache, but we show them as a limit study.
The performance impact was not significant, since BI from a $\mu$thread
overlapped with execution of other $\mu$threads, hiding the latency. 
In addition, when CXL memory BW is saturated, fetching some data from the host 
through CXL port can provide additional BW for moderate dirty cacheline ratios, countering
the BI latency impact.

\begin{figure}
\centering
\includegraphics[width=1\linewidth]{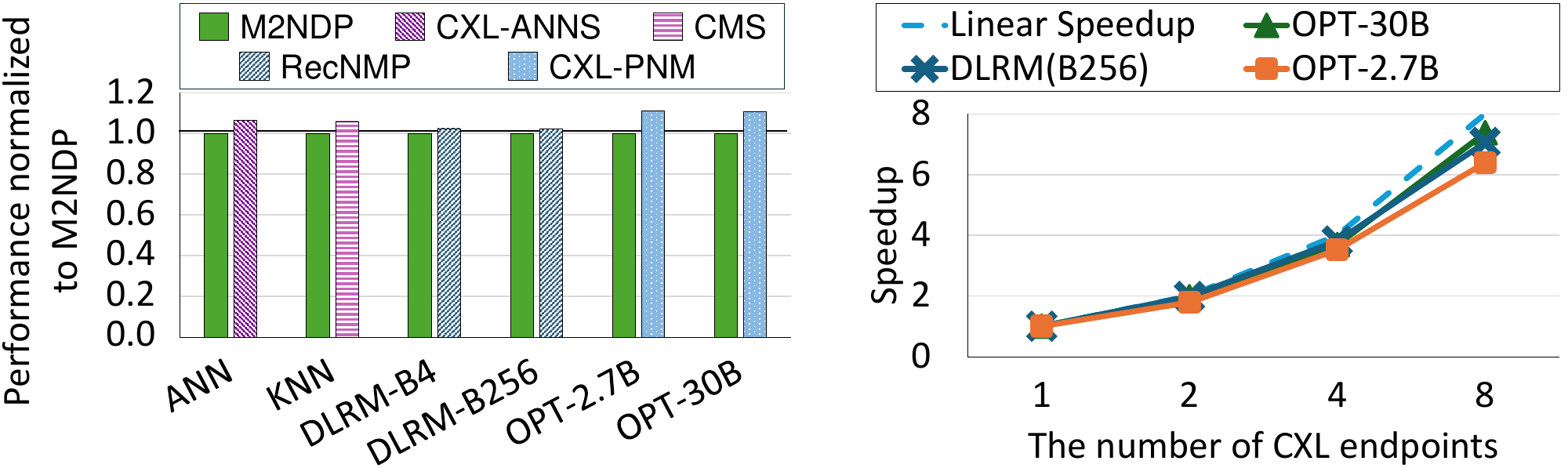}
 \vspace{-.09in}
\hbox{\hspace{0.4in}\footnotesize{(a)} \hspace{1.5in} \footnotesize{(b)}\hspace{0.1in}}
\caption{(a) Performance of domain-specific
CXL-NDP using PEs from prior works (CXL-ANNS~\cite{cxl-anns}, CMS~\cite{cxl_cms_hynix_cal}, RecNMP~\cite{recnmp}, and CXL-PNM~\cite{cxl-pnm}). For ANN and KNN (from CMS~\cite{cxl_cms_hynix_cal}),
we assumed top-K algorithm is executed on the host, overlapping with NDP~\cite{cxl-anns}. 
We assumed a sufficient number of PEs to saturate the memory BW. (b) Scalability with \mmndp{}-enabled
CXL switch with varying number of passive CXL memories.}
\label{fig:DSA_CXL_switch}
\end{figure}

\noindent
{\bf Comparison to domain-specific NDP.}
Compared to using processing elements (PEs) from prior domain-specific NDP works,
\mmndp{}'s performance was within 6.5\% of their performance on average %
(Fig.~\ref{fig:DSA_CXL_switch}a). 
For the memory-bound workloads, \mmndp{} was able to nearly saturate the memory BW by $\sim$81.6\% 
even with the general-purpose design, although domain-specific PEs sometimes exhibited  
higher row buffer locality and utilized memory BW slightly better.

\noindent
\textbf{Scalability of \mmndp{}-enabled CXL switch.}
Even if \mmndp{} were implemented in a CXL switch (\S \ref{sec:passive_cxl}), the performance scaled well with 
the number of passive CXL memories, achieving 6.39-7.38$\times$ speedups with 8 CXL memories by using multiple CXL ports of the switch (Fig.~\ref{fig:DSA_CXL_switch}b).

\begin{figure}
\centering
\includegraphics[width=1\linewidth]{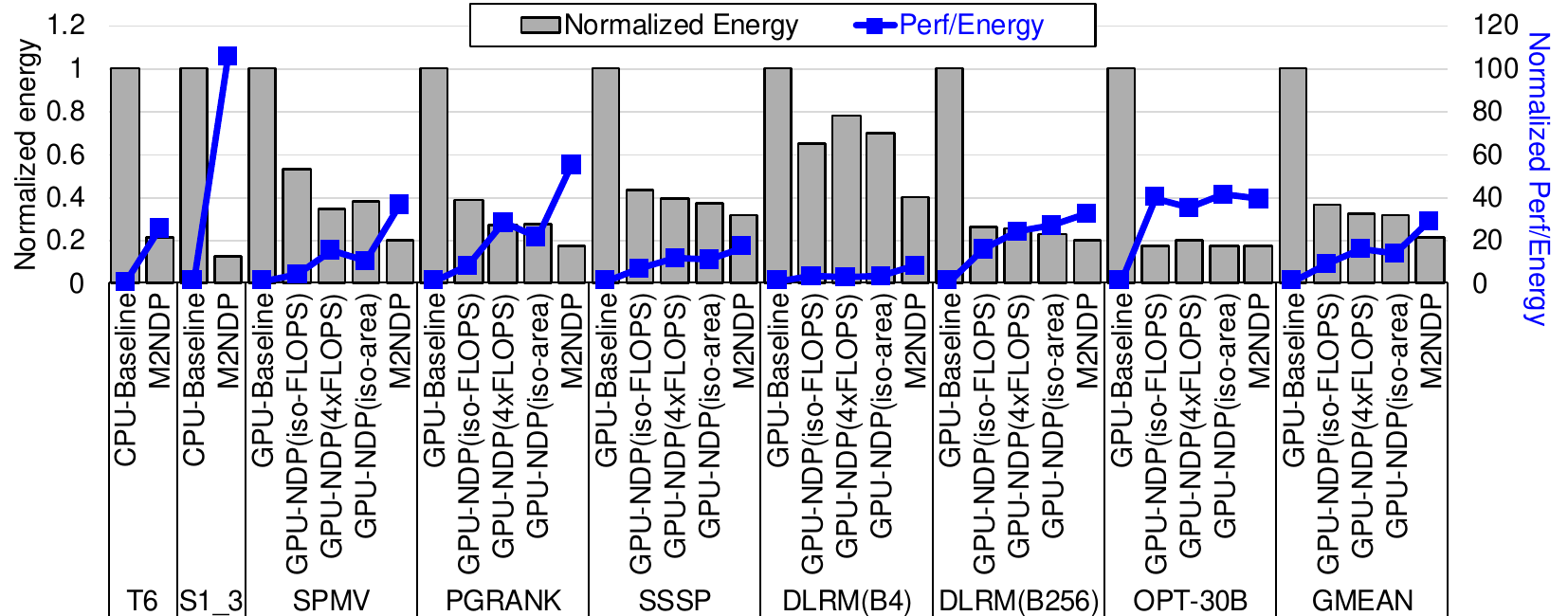}
\caption{
Energy and
performance per energy normalized to baseline CPU and GPU for OLAP and GPU workloads respectively.
\textfw{T6} and \textfw{S1\_3} denote TPC-H Q6 and SSB Q1.3.
GMEAN is calculated for GPU workloads only.}
\label{fig:energy}
\end{figure}

\subsection{Energy}

Compared to the baselines, \mmndp{} significantly improved the performance per energy by up to 106$\times$ and 32.0$\times$ on average (Fig.~\ref{fig:energy}).
For \textfw{OLAP}, \mmndp{} substantially reduced energy consumption by up to 87.9$\%$ 
(83.9$\%$ on average) 
compared to the CPU baseline without NDP by reducing data movement over the CXL link and static/constant 
energy with lower runtime.
Similarly, for GPU workloads, \mmndp{} also significantly reduced energy compared to the baseline by 78.2$\%$ on average.
Compared to the \textfw{\gpundparea{}}, we reduced energy by up to 
76.3\% (31.4\% on average).

\subsection{Hardware Cost}
\label{sec:cost}

We estimated the areas of caches and TLBs in the NDP unit using CACTI 6.5 and scaled them to 7~nm with node-scaling factor from~\cite{techscaling}.
The area of register files (integer, float, and vector) is estimated to be 0.25~$mm^2$. Each NDP unit has a unified L1 and scratchpad memory of 0.45~$mm^2$. With each $\mu$thread slot occupying 0.002~$mm^2$, a single NDP unit with compute units from \cite{fpnew} occupies 0.83~$mm^2$. Thus, the 32 NDP units that we assumed in the evaluation are estimated to incur an area overhead of only 26.4~$mm^2$.%

\section{Related Work}
\label{sec:related_work}

\subsection{CXL Memory Expander}

Several works studied the performance impact of CXL memory on cloud 
workloads and proposed memory placement schemes~\cite{tpp, cxl_uiuc, smt_cxl} as well as memory pooling~\cite{pond, cxl_pooling}. DirectCXL~\cite{direct_cxl} also
demonstrated the performance benefits of CXL.mem over RDMA. 
 D. D. Sharma~\cite{cxl_hoti, cxl_scale} analyzed the CXL architecture and its performance. 

\vspace{-.02in}
\subsection{Near-Data Processing and Processing-In-Memory }
\label{sec:ndp_related}
\vspace{-.02in}

\noindent
{\bf NDP in memory expanders.} 
Several recent works proposed application-specific NDP in a memory expander 
or disaggregated memory for genome analysis~\cite{beacon},
recommendation model~\cite{cxl_cms_hynix_aicas, tensor_dimm, tensor_casting},
nearest neighbor~\cite{cxl_cms_hynix_cal, cxl_anns}, and DNN parameter server~\cite{coarse}.
In contrast, we propose a general-purpose NDP architecture for CXL memory
to overcome their limited flexibility.

\noindent
{\bf PIM.} 
Recent DRAM-PIM designs implemented PIM units in DRAM to exploit the high DRAM-internal BW across all banks, targeting DNNs~\cite{hbm_pim_samsung, newton, gddr6-aim}
or other data-parallel workloads~\cite{upmem}.
They have different trade-offs, including memory bandwidth available, 
flexibility (e.g., instructions supported), 
communication between PIM units, and virtual memory support within PIM kernel.
However, PIM reduces memory capacity~\cite{newton} and may not be suitable for workloads
with huge memory footprints~\cite{gpt-3, recssd, graph500, tpc-h}.
PIM can also be combined with NDP in the same CXL memory
for computation that cannot be localized in a single DRAM chip. 

\noindent
{\bf NDP in SSD.}
Several works explored NDP in SSD using CPU cores~\cite{biscuit,lambdaio, nkv} or 
FPGA~\cite{smartssd, ins-dla, rm-ssd, nkv} to exploit the high bandwidth and low latency
available internally.
However, there are significant gaps between DRAM and flash in terms of 
BW (e.g., 10~GB/s within SSD vs. 100s GB/s in CXL memory) and latency (10s of $\mu$s for flash vs. 10s of ns for DRAM).
Still, for workloads with low BW demand (e.g., cold KV stores), NDP in SSD can be useful.
Since our NDP units are memory device-agnostic and can 
saturate DRAM BW while being more cost-effective than CPU or GPU cores,
they can be employed in the SSD for efficient general-purpose NDP.
If CXL is used for the SSD's interface,
our \mmfunc{} can also enable low-overhead kernel offloading. 
The speedup by NDP in SSD would be largely determined by its internal BW.

\noindent
{\bf Other NDP approaches.} 
Application-specific NDP in HMCs has been proposed
for DNNs~\cite{tetris, winograd_ndp_micro18, heteropim}, linked-lists~\cite{pointer-chasing, llt_pact16},
and graph workloads~\cite{tesseract}.
For programmable NDP, FPGA/CGRA has been proposed~\cite{hrl,axdimm, chameleon, recnmp, nda, tako},
but they pose programmability challenges of mapping application algorithms to HW logic. 
Several works proposed placing simple NDP logic for very fine-grained NDP~\cite{tom, ndp_sc, pei}, 
but they do not effectively support coarse-grained NDP and are not suitable for data-intensive NDP because
the large number of offload command packets required can create a link BW bottleneck.
Furthermore, they require modifying the memory protocol.
These approaches also cannot work independently of the host CPU/GPU and are tightly
coupled with the thread on the host -- e.g., they require the host to send input data for each NDP thread.
Some prior works introduced CPU or GPU cores in 
HMCs~\cite{mondrian, ndc, heteropim, top_pim},
but our proposed \mmuthr{} can achieve higher efficiency 
with lightweight $\mu$threads and flexible
utilization resources (\S\ref{sec:uthreading} and \ref{sec:perf}).
Several works explored offloading NDP operations to buffer chips of 
DIMMs~\cite{netdimm, memory-channel-network, abc-dimm, dimm-link}. They are orthogonal
to \mmndp{} and can be used in the DIMMs of CXL memory.

\section{Conclusion}

In this work, we propose memory-mapped NDP (\mmndp{}) which enables a 
cost-effective, general-purpose NDP in CXL memory expanders
by combining memory-mapped function (\mmfunc{}) and memory-mapped $\mu$threading (\mmuthr{}).
\mmfunc{} leverages the unmodified CXL.mem protocol for lightweight communication
between the host and CXL device. It facilitates efficient 
NDP kernel launch and management, 
avoiding the high overhead of traditional PCIe/CXL.io-based schemes.
\mmuthr{} introduces $\mu$thread, a lightweight thread with minimal register allocation,
allowing a sufficient number of $\mu$threads to be concurrently executed on a low-cost NDP unit.
Allocation/deallocation of NDP unit's resources including $\mu$thread slots are also
done more flexibly compared to GPU SMs, achieving higher resource utilization. 
Directly mapping $\mu$threads to memory and providing scalar units also 
address the overhead of SIMT-only GPU warps.
Compared to the baseline host processor with a passive CXL memory expander,
\mmndp{} can achieve significant speedups (up to 128$\times$) while reducing
energy (by up to 87.9\%) for various applications 
that require large memory capacity,
including in-memory OLAP, KVStore, LLM, DLRM, and graph analytics.

\section*{Acknowledgment}
This work supported in part by the MSIT (Ministry of Science and ICT), Korea,
under the ITRC (Information Technology Research Center) support program 
(IITP\_2024-RS-2024-00437866) supervised by the IITP (Institute of Information \& communications Technology Planning \& evaluation),
IITP grants (No. RS-2021-II210871, No. RS-2021-II210310, and No. RS-2019-II191906),
and National Research Foundation of Korea (NRF) grants 
(No. RS-2023-00277080 and No. RS-2024-00415602).
This work was also partially supported by SK hynix.
We would like to thank the anonymous reviewers for their insightful comments.
Gwangsun Kim is the corresponding author.

\balance
\bibliographystyle{IEEEtranS}
\bibliography{references}

\end{document}